\documentclass[twocolumn]{article}          
\usepackage[left=2.00cm, right=2.00cm, top=2.00cm, bottom=2.00cm]{geometry}
\usepackage{hyperref}
\usepackage{amsthm}    

\usepackage{subfigure}
\usepackage[ruled,linesnumbered]{algorithm2e}

\usepackage{amsmath}
\usepackage{tabularx}
\usepackage[all]{xy}
\usepackage{graphicx}
\usepackage{txfonts}
\usepackage{color}

\graphicspath{{./figures/}}

\newtheorem{definition}{Definition}     
\newtheorem{theorem}{Theorem}       
\newtheorem{proposition}{Proposition}  

\newcommand{\E}{\mathbf{E}}     
\newcommand{\argmin}{\operatornamewithlimits{argmin}}
\newcommand{\vectr}[1]{\boldsymbol{#1}}  
\newcommand{\pnt}[1]{\boldsymbol #1}    
\newcommand{\vpi}{\vectr{\pi}}   
\newcommand{\cald}{\mathcal{D}}    
\newcommand{\calx}{\mathcal{X}}  
\newcommand{\calv}{\mathcal{V}}   
\newcommand{\distset}{\Omega}  

\newcommand{\loss}{\mathcal{L}}  
\newcommand{\eloss}{\Psi}     

\newcommand{\calr}{\mathcal{R}}   
\newcommand{\calk}{\mathcal{K}}   
\newcommand{\calf}{\mathcal{F}}    
\newcommand{\calc}{\mathcal{C}}  
\newcommand{\calp}{\mathcal{P}}    
\newcommand{\caln}{\mathcal{N}}    

\newcommand{\pv}{\mathbf{vec}}

\newcommand{\mt}[1]{\color{black} #1 \color{black}}

\newcounter{ncomm}

\usepackage{authblk}       
\usepackage{varwidth}

\title{Optimal noise functions for location privacy on continuous regions
\thanks{The final publication (in the International Journal of Information Security) is available at Springer via \url{https://doi.org/10.1007/s10207-017-0384-y}}}

\author[1,2]{Ehab ElSalamouny}
\author[3]{S\'ebastien Gambs}
\affil[1]{\protect\begin{varwidth}[t]{\linewidth}\protect\centering INRIA, France \protect\end{varwidth}}
\affil[2]{\protect\begin{varwidth}[t]{\linewidth}\protect\centering Faculty of Computers and Informatics, Suez Canal University, Egypt \protect\end{varwidth}} 
\affil[3]{Universit\'e du Qu\'ebec \`a Montr\'eal (UQAM), Canada}
\date{}

\begin{document}
\maketitle

\begin{abstract}
Users of location-based services (LBSs) are highly vulnerable to privacy risks since they need to disclose, at least partially, their locations to benefit from these services. 
One possibility to limit these risks is to obfuscate the location of a user by adding random noise drawn from a noise function. In this paper, we require the noise functions to satisfy a generic location privacy notion called 
$\ell$-privacy,
which makes the position of the user in a given region $\calx$ relatively indistinguishable from other points in $\calx$.
We also aim at minimizing the loss in the service utility due to such obfuscation.  
While existing optimization frameworks regard the region $\calx$ restrictively as a finite set of points, we consider the more realistic case in which the region is rather \emph{continuous} with a non-zero area. 
In this situation, we demonstrate that circular noise functions are enough to satisfy $\ell$-privacy on $\calx$ and equivalently on the entire space without any penalty in the utility. 
Afterwards, we describe a large parametric space of noise functions that satisfy $\ell$-privacy on $\calx$, and show that this space has always an optimal member, regardless of $\ell$ and $\calx$. 
We also investigate the recent notion of $\epsilon$-geo-indistinguishability 
as an instance of $\ell$-privacy, and prove in this case that with respect to any increasing loss function, the planar Laplace noise function is \emph{optimal} for \emph{any} region having a nonzero area.
\end{abstract}

\section{Introduction}

The popularity of hand-held devices, such as smartphones, that have positioning capabilities has lead to the development of Loc\-ation-Based services (LBSs). 
In an LBS, the device of a user sends a request together with his geographical position to the service provider who personalizes the service according to the reported location. 
The usefulness of these LBSs comes at the cost of various privacy risks as discussed by \cite{Krumm:2007,Golle:2009,Freudiger:2012}. 
For example, based on the disclosed locations of the user, an adversary can identify the points of interests of a user, such as the home and workplace, predict his mobility and even reconstruct part of his social network.    

To limit these risks, one possibility for achieving location privacy is to make the position of a user indistinguishable to some degree from other locations. 
A recent trend of research \cite{Shokri:2011:QLP,Shokri:2011:QLP:CSLE,Andres:2013:indist,ehab:2016:l-privacy} has been directed to obfuscating the user's location in the submitted queries and has lead to several quantifications of location privacy. 
For instance, the authors of \cite{Shokri:2011:QLP,Shokri:2011:QLP:CSLE} have developed a framework in which the location privacy of the user is measured by the expected adversary's error in estimating the user's real location. 
However, this quantification depends on the user's prior distribution (\emph{i.e.}, his probabilities to be in the individual points of the considered space) and also on the strong assumption that the adversary knows this prior.  

Since it is hard to control or even to assess the knowledge of the adversary, another work \cite{Andres:2013:indist} has introduced the notion of $\epsilon$-geo-indistin\-guishability, which abstracts away from both the knowledge of the adversary 
and the prior of the user.
This notion describes the required protection as a guarantee on the obfuscation mechanism itself. 
Informally, a mechanism $\calk$ should not report an output that influences too much the knowledge of the adversary about 
the user's real location. 
More precisely, a mechanism $\calk$ satisfies $\epsilon$-geo-indistinguish\-ability if the log of the ratio between the probability of reporting an output when the user is at location $\pnt i$, and that probability when he is instead at location $\pnt j$ does not exceed a distinguishability $\epsilon\, d$ in which $\epsilon>0$ is a fixed privacy parameter and $d$ is the distance between $\pnt i$ and $\pnt j$. 
This means that the user's position is hardly distinguishable from nearby points, while being increasingly (\emph{i.e.}, at a linear rate) distinguishable from far away points.  
The notion of $\epsilon$-geo-indistinguishability is inspired from differential privacy, which was proposed in \cite{Dwork:06:ICALP} to protect the privacy of the participants in statistical databases. 
In principle, the addition or removal of a participant in the database should have a minor impact on the output of algorithm operating on the database. 
In that sense, $\epsilon$-geo-indisting\-uishability, similarly to differential privacy, abstracts from the adversary's knowledge, and restricts the information disclosed through the mechanism to the observer. 

The idea of restricting the distinguishability between each pair of locations in a geographical region $\calx$ is generalized in \cite{ehab:2016:l-privacy} to give rise to the notion of $\ell$-privacy. 
Here $\ell(.)$ is a function that specifies for every distance a maximum level of distinguishability. 
The function $\ell(.)$ can take various forms depending on the user's privacy requirements. 
For example, if the distinguishability between two points is required to increase linearly, setting $\ell(d)= \epsilon\, d$ yields $\epsilon$-geo-indisting\-uishability. 
Alternatively, if only the distinguishability between nearby points (within distance $D$) is required to be restricted, setting $\ell(d)=\{\epsilon\,\, \mbox{if}\,\, d\leq D,\,\, \mbox{and}\,\,\infty\,\,\mbox{otherwise}\}$, leads to another instance called ($D$, $\epsilon$)-location privacy \cite{ehab:2016:l-privacy}. 

Obfuscating the position reported to the LBS provider causes a degradation in the quality of the obtained service since it is tuned to the reported location instead of the real one. 
This degradation is typically measured by a \emph{loss} function $\loss(d)$ specifying the loss (as a non-negative number) when the distance between the real position of the user and the reported one is $d$. 
The utility of the mechanism for a user is therefore measured by the expected value of the loss function, taking into account the prior distribution of the user and the probabilistic obfuscation performed by the mechanism. 

In this work, our main objective is to provide a mathematically grounded framework that allows to optimize the trade-off between the utility of the LBS requested by the user and his location privacy within a geographical region. 
A previous approach that was adopted in \cite{Bordenabe:2014:CCS} for the case of $\epsilon$-geo-indisting\-uishability is to regard the region as a finite set of points $\calx$ and to assume that the outputs of a mechanism are also drawn from $\calx$. 
%
In this situation, an optimal mechanism is obtained by solving a linear optimization problem that minimizes the expected loss (taking user's prior into account) subject to the privacy constraints. 
Here, the main difficulty is that the number of linear constraints is too large because of the restriction of the distinguishability between every two points in $\calx$, and considering also every output of the mechanism. 
Despite the improvement proposed by the authors of \cite{Bordenabe:2014:CCS} to reduce the number of constraints, the size of $\calx$ has still to be very small (\emph{e.g.}, 50 to 75 points) to solve the problem in a reasonable time. 
%
%

\begin{figure*}[t]
	\centering
	\subfigure[]{
			\includegraphics[height=2.60in]{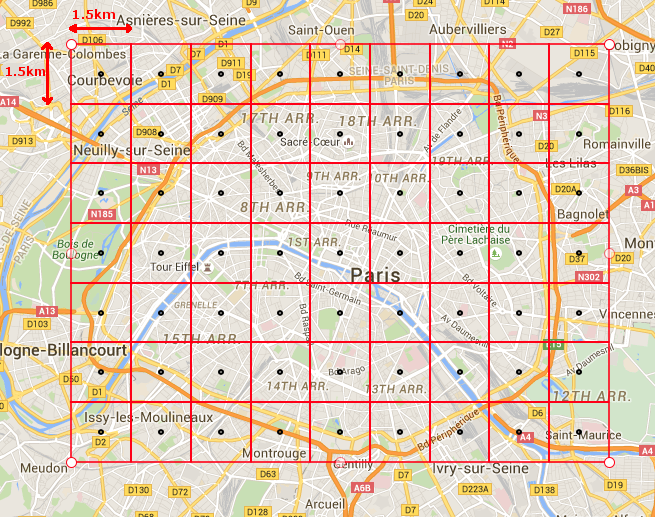}
			\label{fig:grid}
	}
	\subfigure[]{
			\includegraphics[height=2.60in]{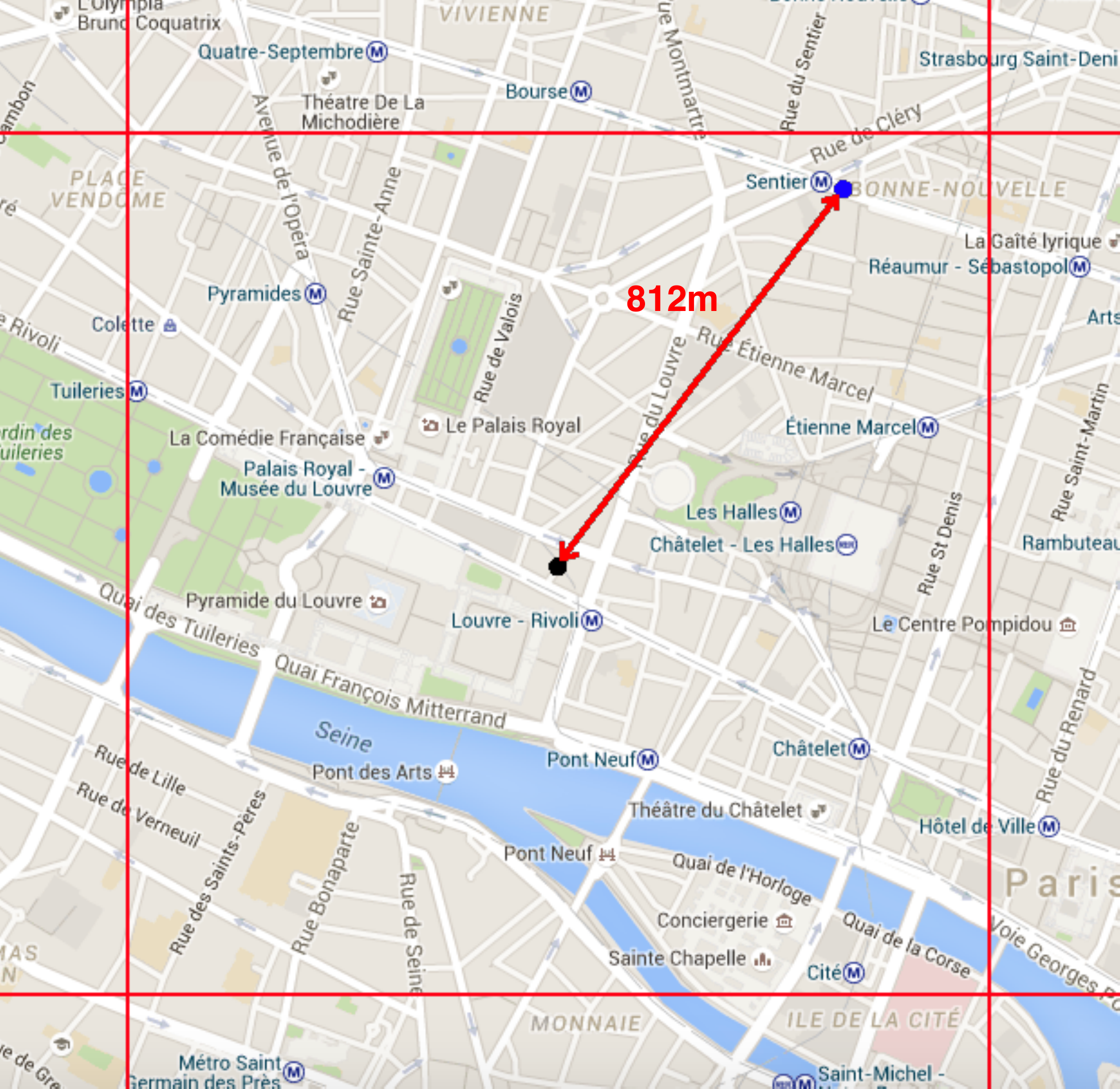}
			\label{fig:large-cells-problem}
	}
	\caption{Approach in which Paris is represented by a finite set of cells: (a) Division of the city into 63 squared cells. 
	              The side length of every cell is 1.5km. (b) One cell in which the user is 812 meters away from the center.}
	\label{fig:paris-grid}
\end{figure*}

While it is always possible to discretize any geographical region into a finite set of points, this discretization
usually incurs a significant loss of quality for the users. 
For example, to construct a mechanism that satisfies $\ell$-privacy for the users in Paris using the above linear optimization, we would need to divide its map into a grid of a feasible size (\emph{e.g.}, 63 cells as shown in Figure \ref{fig:grid}), making every cell  1.5km $\times$ 1.5km. 
In this discretization scheme, the position of every user is always approximated by the center of the enclosing cell before being obfuscated by the mechanism. 
\mt{
Figure \ref{fig:large-cells-problem} displays one cell in which a user located near its north-east corner asks for the nearest restaurant to his position. 
In this case, he would get an answer that is tailored, in the best case, to the center of his cell, which is 0.812km away from him. 
}
It is clear that the situation gets more problematic as we consider larger regions.
      
We take a different approach centered on mechanisms that we call ``symmetric''. 
In these mechanisms, a single distribution $\calp$, called the ``noise distribution'' is used to sample the noise added to the user's location to produce the reported output. 
Since the added noise is essentially an Euclidean vector, the distribution $\calp$ is also regarded as a probability measure on the 
subsets of the Euclidean vector space $\mathbb{E}^2$.
This distribution can be described more succinctly in many situations by a probability density function (pdf) $\calf$, which we refer to as the ``noise function'' of $\calp$.  
%
%
This scheme is both simple and scalable with respect to the topology and the size of the considered region $\calx$ since it is based on one probability distribution (\emph{i.e}, on the noise) that is used at every position of the user in $\calx$. 
Moreover, the expected loss is independent of the user's prior, making the notion of an optimal noise dependent only on the region $\calx$ and the considered loss function.  
In this work, we provide a framework that investigates the above approach in the general setting of the distinguishability (privacy) function $\ell(.)$ and the region of interest $\calx$, aiming to find the optimal noise function with respect to an arbitrary loss function. 
More precisely, our main contributions can be summarized as follows. 

\mt{
\paragraph{Main contributions.} 
\begin{itemize}
\item
We extend symmetric mechanisms \cite{ehab:2016:l-privacy} by using their noise distributions instead of their pdfs (\emph{i.e.}, noise functions) since these latter ones may not exist in some cases (\emph{e.g.}, when the distribution assigns non-zero probabilities to discrete vectors). 
In this extension, we describe the precise condition on a distribution $\calp$ to exhibit a noise function, and the condition on this function to satisfy $\ell$-privacy. 
This privacy condition turns to be independent of the continuity restriction that was imposed in \cite{ehab:2016:l-privacy} on all noise functions. 
\item When the region $\calx$ is continuous with a non-zero area, we prove that some practical instances of $\ell$-privacy are satisfied on $\calx$ \emph{only if} they are satisfied on the entire space $\mathbb{R}^2$. 
Based on this result, the class of circular noise functions turns to be general enough (\emph{i.e.}, without any penalty in the utility due to restriction to this class) to satisfy $\ell$-privacy on \emph{any region} having a non-zero area. 
This extends the special case in which the region is a disc in $\mathbb{R}^2$ as shown in \cite{ehab:2016:l-privacy}.  
%
\item For any setting of distinguishability function $\ell$, set of locations $\calx$ and loss function $\loss$, we describe precise conditions that allow a space of noise functions to have an optimal member for $\calx$ with respect to $\ell$ and $\loss$. 
Based on these conditions, we describe a parametric space of noise functions that \emph{always} admits such an optimal member.
%
\item We consider the instance $\ell(d)= \epsilon\, d$, which corresponds to the notion of $\epsilon$-geo-indistinguishability \cite{Andres:2013:indist}, and prove that in this setting the planar Laplacian noise function (a two-dimensional version of the Laplace density function) is optimal, with respect to \emph{any} increasing loss function and for \emph{any} region $\calx$ having a non-zero area. 
\end{itemize} 
}

\paragraph{Outline of the paper.} First in Section \ref{sec:related-work}, we review the related work before introducing in Section \ref{sec:preliminaries} some preliminaries, such as the notions of mechanisms, $\ell$-privacy and the utility measure. 
Then, in Section \ref{sec:noise-distributions} we develop the formal tools to analyze the privacy of noise distributions and their corresponding noise functions. 
Afterwards in Section \ref{sec:cont-regions}, we focus on continuous regions having nonzero areas and discuss the conditions of satisfying $\ell$-privacy on them before discussing in Section \ref{sec:optimal-noise} the existence of optimal noise functions considering an arbitrary setting of the distinguishability function $\ell$, the region $\calx$ and the loss function $\loss$. 
As a case study, we describe in Section \ref{sec:geo-ind} the optimal noise function for $\epsilon$-geo-indistinguishability and finally summarize our conclusions and directions for future work in Section \ref{sec:conclusions}.

\section{Related work}
\label{sec:related-work}

A possibility to define location privacy is with respect to the ability of an adversary to identify the user's location \cite{Shokri:2010:unraveling}. 
One of the first attempts to achieve location privacy in this direction was to hide the association between the user's identity and his location by removing his identity from the request submitted to the LBS provider or replace it with a pseudonym \cite{Pfitzmann:2001:AUP}. 
However, it turns out that the user's identity can be uncovered by correlating his disclosed locations with some 
background knowledge \cite{Beresford:2003:LPP,Hoh:2006:ESP-Traffic,DBLP:journals/jcss/GambsKC14}.
This issue motivated recent approaches focusing on obfuscating the user's location itself before sending it to the 
server. 
For example, the authors of \cite{Gruteser:2003:STC,Gedik:2005:LPM} proposed a $k$-anonymization of the user location, in the sense that the region reported to the LBS provider, is called a ``cloak'', and ensures that the user is indistinguishable from $k-1$ other users. 
However, as shown by \cite{Shokri:2010:unraveling}, this guarantee may be sometimes inconsistent with the location privacy of the requesting user, for instance if $k$ users are in the same location or at least in a small area. 
In addition, the protection provided by this ``cloaking'' technique depends heavily on the 
background knowledge of the adversary.
To address this shortcoming, the authors of  \cite{Shokri:2011:QLP,Shokri:2011:QLP:CSLE} have developed another metric for location privacy, which is the expected error of the adversary's estimation of the user's location. 
The larger this error is, the higher level of privacy is given to the user. 
In this quantification, it is explicitly assumed that the adversary knows the user's prior. 

Since it is hard in practice to assess the knowledge of adversaries, specially in the existence of public sources of information \cite{ehab:2016:l-privacy}, a recent concept that is inspired from differential privacy \cite{Dwork:06:ICALP} is 
to quantify location privacy instead by the amount of information leaked through the privacy mechanism itself.
Therefore, this makes this measure independent of both the user's prior and the adversary's knowledge. 
\mt{
Differential privacy has been used for instance by the authors of \cite{Chen:12:DPT} in a non-interactive setting to sanitize the transit data of the users of Montr\'eal transportation system. To allow such sanitization despite the inherent high-dimensiona\-lity of the considered data, the authors adopted a data-dependent approach to restrict the output domain of the sanitization mechanism in the light of the underlying database. 
In our work, we focus on interactive mechanisms sanitizing the user's location each time he sends a request to an LBS. 
An adaptation of differential privacy in this setting was 
%
%
proposed by the authors of \cite{Andres:2013:indist} in which the distinguishability between the user's location and another point (in a fixed domain $\calx$) increases linearly with the distance between the two points.
This makes the user's location indistinguishable from nearby points, while being increasingly distinguishable from further away points.
} 
A generalization of this model has been proposed in \cite{ehab:2016:l-privacy} in which the distinguishability, modeled by a generic function $\ell(.)$, between two points  still depends on the distance between them, but may take various forms depending on the privacy requirements of the user. 
The article \cite{ehab:2016:l-privacy} introduced also a restricted form of ``symmetric mechanisms'', which we extend in terms of the underlying noise distributions. 

With respect to optimizing the trade-off between privacy and the expected loss, in addition to \cite{Bordenabe:2014:CCS} which we already mentioned in the introduction, the authors of \cite{Shokri:2012:PLP} considered this problem from a different perspective. They relied on the view of location privacy as the expected adversary's error in estimating the user's real location (as in \cite{Shokri:2011:QLP,Shokri:2011:QLP:CSLE} above) and proposed to construct the mechanism that maximizes the user's privacy, while respecting a certain threshold on the utility.
They also assume that the adversary has an optimal strategy that exploits his knowledge about the user's prior to guess the real location. 
This construction is performed by solving a linear optimization problem in which the number of constraints is quadratic with 
respect to the number of locations in the considered region $\calx$, and therefore has the same efficiency limitations 
of the methodology used in \cite{Bordenabe:2014:CCS}. 

\mt{
According to the distinction made by \cite{Shokri:2011:QLP:CSLE} between \emph{sporadic} and \emph{continuous} location exposure, we focus in this article on the sporadic case in which the locations reported by the user are sparsely distributed over time such that they can be considered independent of each other. 
In this case it is sufficient to sanitize each single location in an independent manner. 
However, in the continuous exposure scenario, the successive reported locations are correlated and therefore other approaches are required to protect the user's entire \emph{trace}. 
For instance, \cite{Chen:12:DPT} describes an efficient mechanism to sanitize a collection of mobility traces in a non-interactive fashion, while in the interactive setting of accessing LBSs, other techniques such as the predictive mechanism \cite{predictive_mechanism} may be used to mitigate the impact of the correlation between the user's successive locations on his privacy.  

Finally, we want to point out that our notion of symmetric mechanism is similar to the noise-adding mechanism of \cite{Geng:OPT_DP:2016} in the sense that both of them add continuous obfuscation noise independently of the original data, and the two articles aim to optimize the added noise. 
However, they differ in two main aspects. 
First, while the mechanism in \cite{Geng:OPT_DP:2016} adds real-valued noise to the numerical query results, our mechanisms add vector-valued noise to the the user's real position. 
Second, while \cite{Geng:OPT_DP:2016} aims to satisfy the standard $\epsilon$-differential privacy for statistical databases, our goal is more general in the sense that we want to satisfy $\ell$-privacy for the user's locations. 
The same authors of \cite{Geng:OPT_DP:2016} described also in another work \cite{Geng:OPT_APPROX_DP:2016} a (nearly) optimal noise-adding mechanism satisfying the approximate $(\epsilon, \delta)$-differential privacy for integer-valued 
and histogram queries.
}

\section{Preliminaries}\label{sec:preliminaries}

We consider a user who may be located anywhere in a certain domain of locations $\calx \subseteq \mathbb{R}^2$, and uses an obfuscation mechanism to produce a noisy position, which is reported to the LBS server. 
Thus, a mechanism is modeled by a probabilistic function $\calk:\calx \to \mathbb{R}^2$ that takes the user's real location $\pnt i \in \calx$ and reports a position $\pnt p \in \mathbb{R}^2$ to the LBS provider. 
We write this probabilistic event as $\calk(\pnt i)=\pnt p$. 
The difference between the reported and real locations is an Euclidean vector $\vec\mu \in \mathbb{E}^2$, which we coin 
as the noise vector, $\vec\mu = \pnt p -\pnt i$
\footnote{
Throughout this paper, we denote the space of points (\emph{i.e.}, locations) by $\mathbb{R}^2$, while the space of Euclidean vectors is represented by $\mathbb{E}^2$.}. 
The input domain $\calx$ of the mechanism is arbitrary and is usually specified to capture all the points that the user may visit. 
The output domain of the mechanism, on the other side, is assumed to be the entire space $\mathbb{R}^2$. 

\subsection{$\ell$-privacy}

A mechanism $\calk$ satisfies $\ell$-privacy (for a user) on a domain of locations $\calx$ if it guarantees that for each region $S \subseteq \mathbb{R}^2$, the probability of reporting a point in $S$ when the user is at $\pnt i$, \emph{i.e.} $P(\calk(\pnt i) \in S )$, is not ``too different'' from that probability when he is instead at $\pnt j$ (both $\pnt i, \pnt j$ are in $\calx$). 
The restriction on this difference between the two probabilities depends on the distance between $\pnt i$ and $\pnt j$ (\emph{i.e.} $|\pnt i - \pnt j|$) and the specification of a distinguishability function $\ell$. 
More formally, we recall the definition of this notion from \cite{ehab:2016:l-privacy}.  

\begin{definition}[$\ell$-privacy~\cite{ehab:2016:l-privacy}]\label{def:l-privacy}
For a distinguishability function $\ell:[0, \infty) \to [0, \infty)$, a mechanism $\calk$ satisfies 
$\ell$-privacy on $\calx$ if for all $S \subseteq \mathbb{R}^2$ it holds 
\[
P(\calk(\pnt i) \in S ) \leq e^{\ell(|\pnt i - \pnt j|)} P(\calk(\pnt j) \in S) 
  \quad \forall \pnt i, \pnt j \in \calx. 
\]
\end{definition}

Note that the level of privacy is controlled by the behavior of $\ell(.)$ with respect to the distance $d$ between the two points. 
The distinguishability function $\ell(d)$ may be also seen as modeling the risk of distinguishing the user's location from others at distance $d$. 
For example, the risk level may get lower as the distance $d$ grows and is accordingly modeled by an increasing $\ell(d)$. 

\subsection{Symmetric mechanisms}
\label{sec:sym}

A mechanism $\calk$ is called `symmetric' if sampling the noise vector is 
independent of the real location of the user \cite{ehab:2016:l-privacy}. More precisely, a symmetric 
mechanism samples a noise vector $\vec\mu$ using a fixed probability distribution $\calp$ on the 
subsets of the vector space $\mathbb{E}^2$ and then reports to the LBS server the user's location after 
adding $\vec\mu$ to it. We call $\calp$ the noise distribution of $\calk$. 

In \cite{ehab:2016:l-privacy}, a symmetric mechanism was defined using the probability density function (pdf) of the distribution $\calp$, assuming that this pdf exists for $\calp$. 
Moreover, this pdf, which is also called a noise function, was assumed to be continuous everywhere in each bounded subregion of $\mathbb{E}^2$, except possibly on finitely many analytic curves. 
In our reasoning about optimality, we will abstract from these assumptions and base our analysis on the noise distribution $\calp$ as a probability measure before studying its pdf (if it exists). 
More precisely in Section \ref{sec:noise-distributions}, we will redefine a symmetric mechanism in a more generic manner using its noise distribution $\calp$, demonstrate the precise conditions on $\calp$ to satisfy $\ell$-privacy, and then proceed to study 
its corresponding pdf (\emph{i.e.}, its noise function).  

\subsection{Loss functions and the expected loss}

The utility of a mechanism for the user is measured by the expected (average) ``loss'' incurred due to reporting noisy locations instead of the real ones. 
This requires specifying a loss function $\loss:[0, \infty) \to [0, \infty)$ that assigns to each noise magnitude a loss value. 
In general, the expected loss depends on the prior probabilities $\vpi$ of visiting the points of $\calx$, and of course on the mechanism. 
However if the mechanism is symmetric (\emph{i.e.}, the noise vector is sampled using a fixed noise distribution $\calp$ as described earlier) the expected loss is independent of $\calx$ and the prior distribution $\vpi$. 
Assuming that $\calp$ has a probability density function (\emph{i.e.}, a noise function) $\calf$, it was shown in \cite{ehab:2016:l-privacy} that the expected loss of $\calf$ with respect to $\loss$ is given by 
\begin{equation}\label{eq:exp-loss-noise}
\eloss(\calf, \loss) = \E[\loss(|\vec\mu|)] = \iint_{\mathbb{E}^2} \, \calf(\vec\mu)\,\loss(|\vec\mu|)\, d \lambda(\vec\mu).  
\end{equation}

In practice, the loss $\loss$ is defined by the user depending on the target LBS. For example, if he wants to query 
the set of nearest restaurants to his position, $\loss$ may be defined as $\loss(x) = x$; \emph{i.e.} the less perturbation of his 
location, the more useful is the response of his query. Alternatively, for a weather forecasting service, $\loss(x)$ may take the 
value 0 if the noise magnitude $x$ is within a certain threshold in which the weather is almost uniform, while it takes larger values
beyond this threshold. 

\section{Noise distributions and noise functions}
\label{sec:noise-distributions}

As mentioned previously in Section \ref{sec:sym}, a symmetric mechanism is determined by its noise distribution $\calp$, which corresponds to its probability measure on the subsets of the vector space $\mathbb{E}^2$. 
Therefore, we define a symmetric mechanism in the following by its corresponding  distribution $\calp$. 

For any set of points $S\subseteq\mathbb{R}^2$, let $\pv(S)$ be the set of position vectors that correspond to the points in $S$.
In addition for any set of vectors $V\subseteq \mathbb{E}^2$, and a vector $\vec u$, let $\tau_{\vec u}(V)$ be the translation image of $V$ by $\vec u$. 
Finally, let $\calp(V)$ be the probability that the sampled noise vector is a member of $V$. 
Then we define a symmetric mechanism using its underlying noise distribution as follows.

\begin{definition}[Symmetric mechanism]\label{def:sym-mech}
A mechanism $\calk$ is said to be symmetric if there is a noise distribution $\calp$ on the subsets of the vector space $\mathbb{E}^2$ such that for every input location $\pnt i$ and a region $S$, it holds that
\[
P(\calk(\pnt i) \in S) = \calp(\tau_{- \pnt i}(\pv(S))). 
\]
\end{definition} 

The above definition means that an output point in $S$ is produced by first sampling a noise vector from $\mathbb{E}^2$ using $\calp$, and then adding this vector to the user's position $\pnt i$. 
It is important to characterize when exactly a noise distribution $\calp$ satisfies $\ell$-privacy on a set of locations $\calx$. 
By Definition \ref{def:l-privacy} of $\ell$-privacy,  the probability of any output of the mechanism should not substantially (subject to the function $\ell(.)$) vary from the probability of this event if the user's position in $\calx$ changes by a vector $\vec u$. 
If a fixed noise distribution $\calp$ is used for sampling noise vectors independently of the input location, this statement can be translated to an equivalent condition on the distribution $\calp$. 
This condition has to take into account all displacements that the user can make in $\calx$. 
Therefore, in the following we denote by $\calv_\calx$ the set of all possible displacement vectors in $\calx$ (\emph{i.e.}, $\calv_\calx = \{ \pnt j  - \pnt i\, :\, \pnt i, \pnt j \in \calx\}$). 

\begin{theorem}[$\ell$-private distributions]\label{thm:l-privacy-dist}
A noise distribution $\calp$ satisfies $\ell$-privacy on the domain $\calx$ if and only if 
\begin{equation}\label{eq:l-privacy-prob}
\calp(V) \leq e^{\ell(|\vec u|)} \calp(\tau_{\vec u}(V)) 
  \quad \forall V \subseteq \mathbb{E}^2, \forall \vec u \in \calv_\calx. 
\end{equation}
\end{theorem}
\begin{proof}
First we show that Def. \ref{def:l-privacy} implies Inequality (\ref{eq:l-privacy-prob}) in the theorem. 
Consider any $V \subseteq \mathbb{E}^2$, and any $\vec u \in \calv_\calx$. Then 
there must be two points $\pnt i, \pnt j \in \calx$ such that $\vec u = \pnt i - \pnt j$. 
Let $S$ be a planar region such that 
$V = \tau_{-\pnt i}(\pv(S))$. 
Therefore 
$\tau_{\pnt i -\pnt j }(V) = \tau_{\pnt i - \pnt j}( \tau_{-\pnt i}(\pv(S))  ) =
\tau_{- \pnt j}(\pv(S))$. 
Using Def. \ref{def:sym-mech} we obtain  
$\calp(V) = P(\calk(\pnt i) \in S)$, and  
$\calp(\tau_{\pnt i -\pnt j }(V)) = P(\calk(\pnt j) \in S)$. 
Now using Def. \ref{def:l-privacy}, we get 
$\calp(V) \leq e^{\ell(|\pnt i - \pnt j|)} \calp(\tau_{\pnt i -\pnt j }(V))$, 
which yields Inequality (\ref{eq:l-privacy-prob}) by substituting $\vec u = \pnt i -\pnt j$. 

Conversely, we show that Inequality (\ref{eq:l-privacy-prob}) implies the inequality in 
Def. \ref{def:l-privacy}. Consider any region $S$, and any $\pnt i, \pnt j \in \calx$. 
Let $V = \tau_{-\pnt i}(\pv(S))$. As shown above, 
$\tau_{\pnt i -\pnt j }(V) = \tau_{- \pnt j}(\pv(S))$. 
Substituting these equalities in (\ref{eq:l-privacy-prob}) with $\vec u = \pnt i - \pnt j \in \calv_\calx$, 
we obtain 
$\calp(\tau_{-\pnt i}(\pv(S))) \leq e^{\ell(|\pnt i - \pnt j|)} \calp(\tau_{- \pnt j}(\pv(S)))$ 
which leads, using Def. \ref{def:sym-mech}, to the inequality of Def. \ref{def:l-privacy}. 
\end{proof}

\subsection{Noise functions}
\label{sec:noise-func}

Since the noise vectors are sampled from the vector space $\mathbb{E}^2$ which is clearly continuous, it makes sense to describe a noise distribution $\calp$ by a corresponding probability density function (pdf) $\calf:\mathbb{E}^2 \to \mathbb{R}^+$. 
We coin this pdf as the ``noise function'' of $\calp$. 
However, in general, this function may not exist for $\calp$.
For instance, if $\calp$ is a distribution on a discrete set of noise vectors in $\mathbb{E}^2$, then $\calp$ has no noise function. The necessary and sufficient condition on $\calp$ to have a noise function is recognized by the Radon-Nikodym theorem \cite[Theorem 5.4]{salamon:2016:integration}, 
which is formulated using the Lebesgue measure $\lambda(V)$ of every subset $V$ of $\mathbb{E}^2$. 
Precisely, a distribution $\calp$ has a noise function if and only if every null subset of $\mathbb{E}^2$, (\emph{i.e.} having Lebesgue measure zero), has also probability $0$. 
In formal terms, this property means that $\calp(V)=0$ whenever $\lambda(V)=0$. 
A distribution $\calp$ that has this property is said to be ``absolutely continuous'' with respect to $\lambda$, and is written as $\calp \ll \lambda$.  
%
%
In this case, the Lebesgue differentiation theorem relates the noise distribution $\calp$ to its noise function $\calf$, and leads to the following important characterization of $\ell$-privacy in terms of $\calf$. 

\begin{theorem}[$\ell$-private noise functions]\label{thm:l-privacy-func}
Let $\calp$ be a noise distribution satisfying $\calp \ll \lambda$. Then $\calp$ and 
its noise function $\calf$ satisfy $\ell$-privacy on a domain $\calx$ 
if and only if there is a null set $\caln \subset \mathbb{E}^2$ such that 
for all vectors $\vec v, \vec v' \in \mathbb{E}^2 \setminus \caln$, it holds 
\begin{equation}\label{eq:priv-func}
\calf(\vec v)  \leq e^{\ell(|\vec v - \vec v'|)} \, \calf(\vec v') \quad 
\mbox{whenever}\quad \vec v -\vec v' \in \calv_\calx.  
\end{equation}
\end{theorem}
\begin{proof}
Since $\calp \ll \lambda$, 
it follows by the Radon-Nikodym theorem that there is a noise function 
$\calf$ on the vector space $\mathbb{E}^2$ satisfying 
$\calp(V) = \iint_V \calf(\vec v)\, d \lambda(\vec v)$, 
for every $V\subseteq\mathbb{E}^2$.  
Now Let $B_\delta(\vec v) \subset \mathbb{E}^2$ be a ball of radius 
$\delta$ around $\vec v$. It follows by the Lebesgue differentiation 
theorem that
\begin{equation}\label{eq:Leb-limit}
\lim_{\delta \to 0}  \calp(B_\delta(\vec v)) / \lambda(B_\delta(\vec v)) = \calf(\vec v)\quad \mbox{a.e. in}\,\,\mathbb{E}^2. 
\end{equation}
In other words there is a null set 
$\caln$ (empty or has $\lambda(\caln)=0$) such that the above equation is satisfied 
for every $\vec v \in \mathbb{E}^2 \setminus \caln$. 
Now consider any $\vec v, \vec v' \in \mathbb{E}^2 \setminus \caln$ such that 
$\vec v -\vec v' \in \calv_\calx$. Then it also holds that $\vec v' -\vec v \in \calv_\calx$. 
Since $\calp$ satisfies $\ell$-privacy on $\calx$, 
it holds by Theorem \ref{thm:l-privacy-dist} that 
$
\calp(B_\delta(\vec v)) \leq e^{\ell(|\vec v'- \vec v|)}\, \calp(\tau_{\vec v' - \vec v}(B_\delta(\vec v))). 
$ 
Note that $\tau_{\vec v' - \vec v}(B_\delta(\vec v))$ is $B_\delta(\vec v')$ 
and therefore 
$\calp(\tau_{\vec v' - \vec v}(B_\delta(\vec v)))= \calp(B_\delta(\vec v'))$. It is also easy to see that 
$\lambda(B_\delta(\vec v)) = \lambda(B_\delta(\vec v'))$. 
Thus we have 
\[
\calp(B_\delta(\vec v))/\lambda(B_\delta(\vec v)) \leq e^{\ell(|\vec v'- \vec v|)}
\calp(B_\delta(\vec v'))/\lambda(B_\delta(\vec v')).
\]
By taking the limits of the above equation when $\delta \to 0$ and substituting the 
two limits using Equation (\ref{eq:Leb-limit}) we obtain 
$\calf(\vec v)  \leq e^{\ell(|\vec v - \vec v'|)} \, \calf(\vec v')$. 

Conversely, suppose that Inequality (\ref{eq:priv-func}) holds for every 
$\vec v, \vec v' \in \mathbb{E}^2\setminus \caln$ such that $\vec v - \vec v' \in \calv_\calx$. 
Consider any fixed $\vec u \in \calv_\calx$. Then by this inequality, it holds that
$\calf(\vec x) \leq e^{\ell(|\vec u|)} \calf(\vec x + \vec u)$ a.e. in $\mathbb{E}^2$. 
Let $\vec y = \vec x+\vec u$. Then by integrating the latter inequality on any set $V$
we get 
$
\calp(V)= \iint_V \calf(\vec x)\, d \lambda(\vec x) 
\leq e^{\ell(|\vec u|)}\, 
\iint_{\tau_{\vec u}(V)} \calf(\vec y)\, d \lambda(\vec y) = e^{\ell(|\vec u|)}\, \calp(\tau_{\vec u}(V)).
$
\end{proof}

The above theorem is useful to check whether a given noise function $\calf$ satisfies (or not) $\ell$-privacy. 
In fact Condition \ref{eq:priv-func} describes the constraints on the values of $\calf$ to satisfy $\ell$-privacy. 
This actually raises another issue, which is central to the objective of this paper. 
This issue concerns whether these constraints can be used to derive an ``optimal'' noise function. 
In general, the answer is negative because for any $\calf$ satisfying $\ell$-privacy, Condition \ref{eq:priv-func} may be violated for some null set $\caln$ that may be anywhere in $\mathbb{E}^2$. 
In other words, if we want to construct an optimal noise function, then for any $\vec v, \vec v' \in \mathbb{E}^2$ such that $\vec v - \vec v' \in \calv_\calx$, we do not know if the inequality in \ref{eq:priv-func} should hold for the values of $\calf$ at $\vec v, \vec v'$ or not. 
However, the answer to the above question is positive if the values of $\calf$ at the vectors in $\caln$ can be ``regulated'' such that (\ref{eq:priv-func}) holds everywhere in $\mathbb{E}^2$. 
In this case, we would have a strict condition that is satisfied for every pair $\vec v, \vec v'$. 
It turns out that such ``regulation'' is possible if the distinguishability function is regular as we define in the following. 

\begin{definition}[Regular distinguishability functions]\label{def:reg-distinguishability}
A distinguishability function $\ell$ is said to be regular 
if for every $\vec v_1, \vec v_2, \vec v_3 \in \mathbb{E}^2$, it holds that 
$
\ell(|\vec v_1 - \vec v_3|) \leq \ell(|\vec v_1 - \vec v_2|) + \ell(|\vec v_2 - \vec v_3|). 
$ 
\end{definition}

Note that $|\vec v_1 - \vec v_3|$ is a metric on vectors and therefore it respects the well known triangle inequality 
$|\vec v_1 - \vec v_3| \leq |\vec v_1 - \vec v_2| + |\vec v_2 - \vec v_3|$. 
Therefore by Definition \ref{def:reg-distinguishability}, a distinguishability function $\ell$ is regular if the triangle inequality for vectors still holds when $\ell(.)$ is applied to every one of its terms.  
An instance of regular distinguishability functions is obtained when the distinguishability 
is proportional to the above metric (\emph{i.e.}, $\ell(d) = \epsilon\, d$). 
This function describes exactly the notion of $\epsilon$-geo-indistinguishability \cite{Andres:2013:indist}, for which we describe an optimal noise function in Section \ref{sec:opt-geo-ind}. 
In general, for any regular distinguishability function $\ell$, the following theorem confirms that every noise function $\calf$ can be always regulated to satisfy the privacy Condition \ref{eq:priv-func} everywhere in $\mathbb{E}^2$. 

\begin{theorem}[(regulating noise functions]\label{thm:l-privacy-func-reg}
Let $\ell$ be a regular distinguishability function. 
Then for every domain of locations $\calx$ and every noise function $\calf$ 
satisfying $\ell$-privacy on $\calx$, 
there is a noise function $\calf' = \calf$ a.e. in $\mathbb{E}^2$ such that 
for all vectors $\vec v, \vec v' \in \mathbb{E}^2$ it holds 
\begin{equation}\label{eq:eq:priv-func-reg}
\calf'(\vec v)  \leq e^{\ell(|\vec v - \vec v'|)} \, \calf'(\vec v') 
\quad\mbox{whenever}\quad \vec v -\vec v' \in \calv_\calx.  
\end{equation}
\end{theorem}
\begin{proof}
Let $\ell$ be regular, and for any set of locations $\calx$ let $\calf$ be a noise function 
satisfying $\ell$-privacy on $\calx$. According to Theorem \ref{thm:l-privacy-func} 
there is a null set $\caln \subset \mathbb{E}^2$ such 
that for every $\vec x, \vec x' \in \mathbb{E}^2\setminus \caln$ it holds 
\[
\calf(\vec x)  \leq e^{\ell(|\vec x - \vec x'|)} \, \calf(\vec x') \quad 
\mbox{whenever}\quad \vec x -\vec x' \in \calv_\calx. 
\]
Define $\calf'$ as follows. For every $\vec x \in \mathbb{E}^2\setminus\caln$, 
let $\calf'(\vec x) = \calf(\vec x)$, and for every $\vec y \in \caln$ let 
$\calf'(\vec y) = \inf\{\ e^{\ell(|\vec y - \vec x|)}\, \calf(\vec x): \vec x \in \mathbb{E}^2\setminus \caln \}$. 
Note that this infimum exists because $\mathbb{E}^2\setminus \caln$ is nonempty and 
$e^{\ell(|\vec y - \vec x|)}\, \calf(\vec x)$ is lower bounded by $0$. 
Observe also that $\calf'= \calf$ a.e. 
In the following we show that Inequality (\ref{eq:eq:priv-func-reg}) holds for every 
two vectors in $\mathbb{E}^2$. 
First, it is easy to see that 
for all $\vec x, \vec x' \in \mathbb{E}^2\setminus \caln$,  
Inequality (\ref{eq:eq:priv-func-reg}) holds since $\calf'=\calf$ at these vectors. 
Now for every $\vec y \in \caln$ and $\vec x \in \mathbb{E}^2\setminus \caln$, 
it holds by the definition of $\calf'$ that $\calf'(\vec y) \leq e^{\ell(|\vec y - \vec x|)}\, \calf'(\vec x)$. 
Based on the hypothesis that $\ell$ is regular, we also claim for every $\vec y \in \caln$ that  
\begin{equation}\label{eq:sup-less-inf}
\begin{split}
\sup \{e^{-\ell(|\vec y - \vec x|)}\,\calf(\vec x) &: \vec x \in \mathbb{E}^2\setminus \caln \}
\leq  \\
&\inf \{e^{\ell(|\vec y - \vec x|)}\,\calf(\vec x): \vec x \in \mathbb{E}^2\setminus \caln \} 
\end{split}
\end{equation}
%
%
which implies that $e^{-\ell(|\vec y - \vec x|)}\,\calf'(\vec x) \leq \calf'(\vec y)$ for all 
$\vec y \in \caln$ and $\vec x \in \mathbb{E}^2\setminus \caln$. 
Thus we conclude that Inequality (\ref{eq:eq:priv-func-reg}) holds for every 
$\vec y \in \caln$ and $\vec x \in \mathbb{E}^2\setminus \caln$. We prove 
Inequality (\ref{eq:sup-less-inf}) as follows. 
Suppose this inequality does not hold for some $\vec y \in \caln$. 
Then there are $\vec x, \vec x' \in \mathbb{E}^2 \setminus \caln$ such that 
$e^{-\ell(|\vec y - \vec x|)}\,\calf(\vec x) > e^{\ell(|\vec y - \vec x'|)}\,\calf(\vec x')$, 
i.e. 
$\calf(\vec x) > e^{\ell(|\vec y - \vec x'|)+\ell(|\vec y - \vec x|)}\,\calf(\vec x')$. 
Since it also holds that 
$\ell(|\vec y - \vec x'|)+\ell(|\vec y - \vec x|) \geq \ell(|\vec x' - \vec x|)$
because $\ell$ is regular, we obtain 
$\calf(\vec x) > e^{\ell(|\vec x' - \vec x|)}\,\calf(\vec x')$ which contradicts 
with the fact that 
$\calf(\vec x) \leq e^{\ell(|\vec x' - \vec x|)}\,\calf(\vec x')$ 
since $\vec x, \vec x' \in \mathbb{E}^2\setminus \caln$. 

Finally consider any $\vec y, \vec y' \in \caln$. We show that 
$\calf'(\vec y') \leq e^{\ell(|\vec y' - \vec y|)} \calf'(\vec y)$. 
Consider any arbitrary small $\delta>0$. By the definition of $\calf'(\vec y)$, 
there must be $\vec x_\delta \in \mathbb{E}^2 \setminus \caln$ such that 
$\calf'(\vec y) \geq e^{\ell(|\vec y - \vec x_\delta|)}\, \calf(\vec x_\delta) - \delta$.
Recalling that $\calf(\vec x_\delta) = \calf'(\vec x_\delta)$, and using the 
inequality $\calf'(\vec y') \leq e^{\ell(|\vec y' - \vec x_\delta|)}\, \calf'(\vec x_\delta)$ 
which was already proved, we obtain
$\calf'(\vec y') \leq e^{\ell(|\vec y' - \vec x_\delta|) - \ell(|\vec y - \vec x_\delta|)}(\calf'(\vec y)+\delta)$. 
Since $\ell$ is regular, it holds that 
$\ell(|\vec y' - \vec x_\delta|) - \ell(|\vec y - \vec x_\delta|) \leq \ell(|\vec y' - \vec y|)$. 
Therefore 
$\calf'(\vec y') \leq e^{\ell(|\vec y' - \vec y|)}(\calf'(\vec y)+\delta)$ for every $\delta >0$. 
Taking the limits of this inequality as $\delta \to 0$ yields  
$\calf'(\vec y') \leq e^{\ell(|\vec y' - \vec y|)} \, \calf'(\vec y)$. 
%
\end{proof} 

Theorem \ref{thm:l-privacy-func-reg} allows us to assume without loss of generality that the privacy Constraints \ref{eq:eq:priv-func-reg} are satisfied for every pair $\vec v, \vec v' \in \mathbb{E}^2$. 
In fact since $\calf'=\calf$ almost everywhere, the integrals of these two functions are the same on any subset of $\mathbb{E}^2$. This means that $\calf'$ is (similar to $\calf$) a valid pdf and also has the same expected loss of $\calf$. 
As mentioned earlier, this conclusion is useful when we derive the optimal noise function satisfying $\ell$-privacy for some domain $\calx$, because we do not need to consider noise functions in which (\ref{eq:eq:priv-func-reg}) is violated on a null set. 

\subsection{Circular noise functions}
\label{sec:circ-noise} 

A noise function $\calf_\calr$ is called ``circular'' if all noise vectors having the same magnitude are drawn with the same probability density \cite{ehab:2016:l-privacy}. 
This probability density is determined by an underlying function $\calr:[0,\infty) \to \mathbb{R}^+$, which we call the ``radial'' of $\calf_\calr$. 
Thus, for every vector $\vec v$ it holds that $\calf_\calr(\vec v) = \calr(|\vec v|)$. 
In this case, it is easy to express the expected loss of $\calf_\calr$ with respect to a loss function $\loss$ as 

\begin{equation}\label{eq:expected loss-radial}
\eloss(\calf_\calr, \loss) = \int_0^\infty  \calr(r) \, \loss(r) \, 2\pi r \,d r.  
\end{equation}
It is also easy to ensure that $\calf_\calr$ assigns total probability 1 to all vectors in $\mathbb{E}^2$ by the following constraint that we coin as the ``total probability law''. 
\begin{equation}
\int_0^\infty \calr(r)\, 2 \pi r \,d r = 1. \label{eq:prob-law-radial}
\end{equation}
%

We now describe the condition on a circular noise function $\calf_\calr$ to satisfy $\ell$-privacy for a domain $\calx$. 
This condition depends on the set 
$
\distset_\calx = \{ (|\vec v|, |\vec v'|) : \vec v, \vec v' \in \mathbb{E}^2,\, \vec v' -  \vec v \in \calv_\calx \}
$
that captures every two noise magnitudes required to have a restricted distinguishability from each other. 
This distinguishability for a pair of magnitudes $(r, r') \in \distset_\calx$ must ensure that every two vectors $\vec v, \vec v'$ having these magnitudes are properly indistinguishable from each other. 
Therefore the distinguishability for $(r, r')$ is exactly the ``minimal'' distinguishability $\ell_\calx(r, r')$ defined as   
$
\ell_\calx(r, r') = \min \, \{\,\ell(| \vec v - \vec v' |) \,: \vec v, \vec v' \in \mathbb{E}^2, \, r= |\vec v|, r' = |\vec v'|, \, \vec v -\vec v' \in \calv_\calx \}.    
$

\begin{theorem}[$\ell$-privacy of circular noise functions]
\label{thm:l-privacy-circ}
A circular noise function $\calf_\calr$ having a radial $\calr$ satisfies 
$\ell$-privacy on a domain of locations $\calx$ if and only if 
there is a discrete set of noise magnitudes $\caln' \subset [0, \infty)$ 
such that for all $r, r' \in [0, \infty) \setminus \caln'$ it holds 
\begin{equation}\label{eq:priv-rad}
\calr(r) \leq e^{\ell_\calx(r, r')}\, \calr(r') \quad\mbox{whenever}\quad (r, r') \in \distset_\calx.
\end{equation}
\end{theorem}
\begin{proof}
Suppose that $\calf_\calr$ satisfies $\ell$-privacy on $\calx$. 
Then its noise distribution $\calp$ also satisfies it. 
By the circularity of $\calf_\calr$, the probability of any ball of radius $\delta$ around a vector $\vec v \in \mathbb{E}^2$ depends only on the magnitude $r$ of $\vec v$ (and $\delta$) regardless of its direction. 
Let $\calp_{r, \delta}$ denote this probability. 
Now $\calp$ satisfies $\ell$-privacy if and only if it satisfies the condition of Theorem \ref{thm:l-privacy-dist} that can be written for $\calp_{r, \delta}$ as
\[
\calp_{r, \delta} \leq e^{\ell_\calx(r, r')} \calp_{r', \delta}
\]
for all 
$(r, r') \in \distset_\calx = \{ (|\vec v|, |\vec v'|) : \vec v, \vec v' \in \mathbb{E}^2,\, \vec v' -  \vec v \in \calv_\calx \}$
and 
$ 
\ell_\calx(r, r') = \min \, \{\,\ell(| \vec v - \vec v' |) \,: \vec v, \vec v' \in \mathbb{E}^2, \, r= |\vec v|, r' = |\vec v'|, \, \vec v -\vec v' \in \calv_\calx \}.    
$
This condition (according to Theorem \ref{thm:l-privacy-dist}) considers all vectors $\vec v, \vec v'$ such that  $\vec v -\vec v' \in \calv_\calx$ and having the magnitudes $r, r'$ respectively. 
The minimum distinguishability $\ell_\calx(r, r')$ is taken to ensure that the distinguishability between every $\vec v, \vec v'$ is properly upper-bounded by $\ell(| \vec v - \vec v' |)$. 
By the Lebesgue differentiation theorem, the derivative of $\calp$ with respect to the Lebesgue measure $\lambda$ on $\mathbb{E}^2$ exists and is equal to $\calf_\calr$ almost everywhere in $\mathbb{E}^2$. 
By the circularity of $\calf_\calr$, this means that for a discrete set $\caln'$ of magnitudes, it holds for every $r \in [0, \infty)\setminus\caln'$ that $\lim_{\delta \to 0} \calp_{r, \delta}/\pi \delta^2 = \calr(r)$. 
Applying this limit to the two sides of the above inequality, we obtain the condition stated by the theorem. 

Conversely we show that this condition implies that $\calf_\calr$ satisfies $\ell$-privacy as follows. 
For every $\vec v, \vec v'$ such that $\vec v - \vec v' \in \calv_\calx$, there must be $r, r' \in [0, \infty)\setminus\caln'$ such 
that $r = |\vec v|, r' = |\vec v'|$.  
Thus $(r, r') \in \distset_\calx$, hence $\calr(r) \leq e^{\ell_\calx(r, r')} \calr(r')$. 
Since $\ell_\calx(r, r') \leq \ell(|\vec v- \vec v'|)$, we have $\calf_\calr(\vec v) \leq e^{\ell(|\vec v - \vec v'|)}\calf_\calr(\vec v')$. 
Note that this inequality holds for all vectors in $\mathbb{E}^2$ except those having magnitudes in $\caln'$. 
Thus, this inequality holds for all vectors in $\mathbb{E}^2 \setminus \caln$ where $\caln$ is the set composed of the union of the discrete set of circles having their radii in $\caln'$. 
Since $\caln$ is clearly a null set in $\mathbb{E}^2$, it follows from Theorem \ref{thm:l-privacy-func} that $\calf_\calr$ satisfies $\ell$-privacy. 
\end{proof}

The minimal distinguishability $\ell_\calx(r, r')$ depends, by its definition, on $\calx$ and $\ell$. 
For example, if $\calx$ is the entire space of locations $\mathbb{R}^2$, and the distinguishability $\ell(d)$ is increasing with $d$, it is easy to see that  $\ell_\calx(r, r')$ is exactly $\ell(|r- r'|)$. 

Based on Theorem \ref{thm:l-privacy-circ}, the trade-off between the location privacy provided by a noise function and its utility can be observed. 
In particular, if the incurred loss increases with the noise magnitude, then to provide a reasonable utility, the noise function should intuitively assign high probability densities to short noise vectors to reduce the loss. 
However in view of Theorem \ref{thm:l-privacy-circ} if this function is too biased, it may violate $\ell$-privacy. 
Optimizing this trade-off is therefore an interesting issue that we investigate in our work.

Now, we proceed by highlighting an important merit of circular noise functions when the domain $\calx$ is a ``disk'' in the planar space $\mathbb{R}^2$. 
Informally, every noise function $\calf$ satisfying $\ell$-privacy can be replaced by a circular one $\calf_\calr$ that both provides the same utility of $\calf$ and also satisfies $\ell$-privacy. 
While this result was proved in \cite{ehab:2016:l-privacy} under a continuity assumption (on noise functions) described in Section \ref{sec:sym}, the following theorem removes the need for this assumption and establishes that result in general when the distinguishability function is regular. 
Furthermore, this theorem gives a stronger statement about $\calf_\calr$: its radial $\calr$ satisfies the condition (\ref{eq:priv-rad}) of $\ell$-privacy without exceptions on a discrete set of magnitudes. 
In this case, we say that $\calf_\calr$ ``strictly'' satisfies $\ell$-privacy on $\calx$.   

\begin{theorem}[Generality of circular noise functions]\label{thm:circular-noise-suffice-l}
Let $\ell$ be a regular distinguishability function, and $\calx$ be a disk in $\mathbb{R}^2$. 
For every noise function $\calf$ satisfying $\ell$-privacy for $\calx$ and for every loss function $\loss$, there exists a circular noise function $\calf_\calr$ (with a radial $\calr$) such that $\eloss(\calf_\calr , \loss)=\eloss(\calf, \loss)$ and strictly satisfies $\ell$-privacy on $\calx$, which means that 
\[
\calr(r)  \leq e^{\ell_\calx(r,r')} \, \calr(r') \quad\quad \forall (r, r') \in \distset_\calx.    
\]
\end{theorem}

\begin{proof}
Since $\ell$ is regular, it holds by Theorem \ref{thm:l-privacy-func-reg} that for every noise function satisfying $\ell$-privacy there is a noise function $\calf$ that satisfies Inequality (\ref{eq:eq:priv-func-reg}) for every two vectors in $\mathbb{E}^2$. 
Let $\calf_\calr$ be a circular noise function (with a radial $\calr$) defined on $\mathbb{E}^2$ using the polar coordinates $(r, \phi)$ of every vector as  $\calf_\calr(r,\phi) = \calr(r) = {1}/{(2\pi)} \int_0^{2\pi}  \calf(r,\theta) \, d \theta$. 
By this definition $\calr$ satisfies the total probability law (\ref{eq:prob-law-radial}) and is therefore a valid radial. 
It can be also verified that $\eloss(\calf_\calr , \loss)=\eloss(\calf, \loss)$ using Equations (\ref{eq:expected loss-radial}) and (\ref{eq:exp-loss-noise}) (as in the proof of Theorem 15 in \cite{ehab:2016:l-privacy}).  
Finally, using the same argument in the proof of Theorem 23 in \cite{ehab:2016:l-privacy}, it follows that $\calr(r)  \leq e^{\ell_\calx(r,r')} \, \calr(r')$ for all $(r, r') \in \distset_\calx$. 
%
\end{proof}

Finally, an important strength of the approach is that sampling a noise vector from circular functions is very simple compared to sampling from non-circular ones. A generic algorithm for this sampling is described in \cite{ehab:2016:l-privacy}.

\section{Noise distributions on continuous regions}\label{sec:cont-regions}

In Section \ref{sec:noise-distributions}, we have established the conditions for a noise distribution, and its corresponding noise function to satisfy $\ell$-privacy on an arbitrary domain $\calx$. 
In the following, we focus on the case when $\calx$ is a continuous region with a nonzero area such as a country, a city or in general a region that contains a dense set of points of interests. 
In this case we find, under a mild condition on $\calx$ and the distinguishability function $\ell$, that satisfying the conditions of $\ell$-privacy on $\calx$ is actually equivalent to satisfying these conditions more widely on the entire planar space $\mathbb{R}^2$.   

\begin{theorem}[Satisfying $\ell$-privacy for continuous regions]
\label{thm:noise-dist-cont-region}
Let $\ell$ be a distinguishability function satisfying for some distance $d_0>0$ that $\ell(d_0 + d) \geq \ell(d_0) + \ell(d)$ for all $d>0$. 
Let also $\calx$ be any region that contains a disk of diameter $d_0$. 
In this case a noise distribution $\calp$ satisfies $\ell$-privacy on $\calx$ if and only if it satisfies $\ell$-privacy on $\mathbb{R}^2$.
\end{theorem}

\begin{proof}
It is clear that if a noise distribution $\calp$ satisfies $\ell$-privacy on $\mathbb{R}^2$, it must satisfy it on $\calx$ since $\calx\subseteq\mathbb{R}^2$.  

Conversely, suppose that $\calp$ satisfies $\ell$-privacy on $\calx$ and $\ell$ satisfies the stated condition. 
We show in this case that $\calp$ must satisfy $\ell$-privacy for $\mathbb{R}^2$. 
More precisely, we demonstrate that the condition of $\ell$-privacy described by Inequality (\ref{eq:l-privacy-prob}) is satisfied on the domain $\mathbb{R}^2$. 
Observe that $\calv_{\mathbb{R}^2}$ is the entire vector space $\mathbb{E}^2$, and therefore for every two points $\pnt i, \pnt j$, we have $\pnt j - \pnt i \in \calv_{\mathbb{R}^2}$. 
Therefore, we proceed by showing that for any $\pnt i, \pnt j$
\[
\calp(V) \leq e^{\ell(|\pnt j - \pnt i|)} \calp(\tau_{(\pnt j - \pnt i)}(V)) 
\quad \forall V\subseteq \mathbb{E}^2.
\]
If $|\pnt j - \pnt i| \leq d_0$, it is easy to see that $\pnt j - \pnt i \in \calv_\calx$ and therefore the above inequality holds since $\calp$ satisfies $\ell$-privacy on $\calx$. 
If otherwise $|\pnt j - \pnt i| > d_0$, there is a sequence of points $\pnt i_0, \pnt i_1, \dots, \pnt i_{n+1}$ on the line connecting $\pnt i$ and  $\pnt j$ such that $\pnt i_0 = \pnt i$ and $\pnt i_{n+1} = \pnt j$,  and every successive two points are $d_0$ apart, except $\pnt i_0, \pnt i_1$ which are at most $d_0$ apart, 
i.e. $|\pnt i_{k+1} - \pnt i_k|=d_0$ for $k=1,2, \dots, n$, and $|\pnt i_1 - \pnt i_0| \leq d_0$.  
Since $\calp$ satisfies $\ell$-privacy on $\calx$ and $\pnt i_{k+1} - \pnt i_k \in \calv_\calx$, we have 
\[
\calp(V) \leq e^{\ell(|\pnt i_{k+1} - \pnt i_k|)} \calp(\tau_{(\pnt i_{k+1} - \pnt i_k)}(V)), 
\quad \forall V\subseteq \mathbb{E}^2
,0\leq k\leq  n 
\]
which implies that
\[
\calp(V) \leq e^{\sum_{k=0}^n \ell(| \pnt i_{k+1}-\pnt i_k |)} \calp( \tau_{(\sum_{k=0}^n \pnt i_{k+1}-\pnt i_k)}(V)). 
\]
Since $\ell(d)+\ell(d_0) \leq \ell(d_0 + d)$ for all $d>0$, it follows that $\sum_{k=0}^n \ell(| \pnt i_{k+1}-\pnt i_k |) \leq \ell(|\pnt i_{n+1}-\pnt i_0|)$. 
It is also clear that $\sum_{k=0}^n \pnt i_{k+1}-\pnt i_k = \pnt j - \pnt i$.  
%
%
Thus $\calp(V) \leq e^{\ell(| \pnt j-\pnt i |)} \calp(\tau_{(\pnt j - \pnt i)}(V))$. 
%
\end{proof}

The above theorem describes a condition on the distinguishability function $\ell(.)$ that can be informally described as follows. For distances $\geq d_0$, the distinguishability $\ell(.)$ increases by a rate that is higher or at least the same as its rate for distances $<d_0$. 
There are various practical situations in which the risk of distinguishing the location of the user is modeled by a distinguishability function having the above behavior. 
In the following, we give some examples of such scenarios. 

\paragraph{\bf $\epsilon$-geo-indistinguishability \cite{Andres:2013:indist}.}
As we mentioned earlier, $\ell$-privacy is instantiated to the notion of $\epsilon$-geo-indistinguish\-ability if the distinguishability function is defined as $\ell(d) = \epsilon\, d$. 
Observe in this case that any $d_0 >0$ satisfies the condition $\ell(d_0 +d) = \ell(d_0) + \ell(d)$ for all $d>0$. 
Remarkably, this condition is satisfied with \emph{every} non-zero value for $d_0$. This implies by Theorem \ref{thm:noise-dist-cont-region} that satisfying $\epsilon$-geo-indistinguishability for any region with a non-zero area is equivalent to having the same protection on the entire space.    
In Section \ref{sec:opt-geo-ind}, we will describe in more details the intuition of this distinguishability function, and also derive the optimal noise function for it. 

\paragraph{\bf $D$-restricted distinguishability functions.}
Consider a user who requires his location to be indistinguishable from others situated within a certain proximity $D$ from him, while allowing his location to be distinguishable from positions beyond that proximity. 
This requirement corresponds to a family of distinguishability functions agreeing on that $\ell(d)=\infty$ for all $d>D$, while they differ from each other in the specification of $\ell(d)$ for $d \in [0, D]$. 
For every member of this family, the condition of Theorem \ref{thm:noise-dist-cont-region} holds with $d_0=D$. 
In fact, it is easy to see in this case that $\ell(d_0 +d) = \infty > \ell(d_0) + \ell(d)$ for all $d>0$. 

In the following, we show an important consequence of Theorem \ref{thm:noise-dist-cont-region}. 
In fact, it turns out that for any region $\calx$ and distinguishability function $\ell$ that satisfy the conditions of Theorem \ref{thm:noise-dist-cont-region}, the class of circular noise functions, presented in Section \ref{sec:circ-noise}, is general enough to provide the same privacy and utility levels that are provided by other noise functions.   

\subsection{Generality of circular noise functions} 

As mentioned in Section \ref{sec:circ-noise}, circular noise functions display the important feature that noise vectors having the same magnitude have also the same probability density. 
Based on this uniformity, it is shown by Theorem \ref{thm:circular-noise-suffice-l} that when $\ell$ is regular and $\calx$ is a ``disk'', there is always an $\ell$-private circular function achieving the same expected loss incurred by another $\ell$-private (non-circular) one. 
Theorem \ref{thm:noise-dist-cont-region} allows us to strengthen this statement to hold not only for disks, but also for broader regions if the conditions that were stated in that theorem for $\calx$ and $\ell$ are satisfied. 

\begin{theorem}[Generality of circular noise functions on continuous regions]
\label{thm:generality-circ}
Let $\ell$ be a regular distinguishability function satisfying for some distance $d_0>0$ that $\ell(d_0 + d) \geq \ell(d_0) + \ell(d)$ for all $d>0$. 
Let also $\calx$ be any region that contains a disk of diameter $d_0$. 
For every noise function $\calf$ satisfying $\ell$-privacy on $\calx$, there is a circular noise function $\calf_\calr$ that strictly satisfies $\ell$-privacy on $\mathbb{R}^2$ and has the same expected loss as $\calf$.  
\end{theorem}
\begin{proof}
Let $\calf$ be a noise function satisfying $\ell$-privacy on $\calx$. 
Since $\ell$ and $\calx$ satisfy the conditions of Theorem \ref{thm:noise-dist-cont-region}, it follows that $\calf$ 
must also satisfy $\ell$-privacy on the entire space $\mathbb{R}^2$. 
Since $\ell$ is regular, and $\mathbb{R}^2$ is circular (with infinite diameter), it holds by Theorem \ref{thm:circular-noise-suffice-l} that there is a circular function that strictly satisfies $\ell$-privacy for $\mathbb{R}^2$, while having the same expected loss of $\calf$.
\end{proof}

According to Theorem \ref{thm:generality-circ}, if the given conditions on $\ell$ and $\calx$ are satisfied, there is no need to use a complex non-circular noise function to sample noise vectors while satisfying $\ell$-privacy. 
The main reason for this is that there is always a circular function $\calf_\calr$ that satisfies the same privacy requirement without any penalty on the expected loss. 
This circular function guarantees $\ell$-privacy not only on $\calx$ but also on the entire space $\mathbb{R}^2$. 
Moreover, the conditions of $\ell$-privacy, stated by Theorem \ref{thm:l-privacy-circ}, are strictly satisfied (\emph{i.e.} without exceptions for any set of noise magnitudes). 
These conclusions are important in particular for identifying an optimal noise function that satisfies a given distinguishability function. 
As a case study we will consider in Section \ref{sec:opt-geo-ind} the instance $\ell(d) = \epsilon\, d$ corresponding to the notion of $\epsilon$-geo-indistinguishability proposed by the authors of \cite{Andres:2013:indist}, and use the aforementioned results to identify the optimal noise function for this instance. To achieve, we formally define in the following section optimal noise functions and discuss their existence. 
 
\section{Optimal noise functions}
\label{sec:optimal-noise}

Since in general, there are many noise functions satisfying a given instance of $\ell$-privacy on a specific region $\calx$, we are interested to find the ``optimal'' one that maximizes the utility (\emph{i.e.}, minimize the expected value of a specific loss function). 
More precisely, we consider a space $\Omega$ of noise functions that satisfy $\ell$-privacy on $\calx$ and define the optimal members in this space in the following manner. 

\begin{definition}[Optimal members in a space of noise functions]\label{def:opt}
Consider a distinguishability function $\ell$, a domain of locations $\calx$, and a loss function $\loss$. 
Let $\Omega$ be a space of noise functions that satisfy $\ell$-privacy on $\calx$. 
A member $\calf \in \Omega$ is said to be optimal in $\Omega$ for $\calx$ with respect to $\loss$ if $\eloss(\calf, \loss) \leq \eloss(\calf', \loss)$ for every $\calf'\in \Omega$. 
\end{definition}

In principle, it is not always guaranteed that the given space of noise functions includes an optimal member even if this space is non-empty. 
Stated differently, it may happen that for every noise function in this space there is another member that has a lower expected loss without ever reaching an optimal one. 
In the following, we address this issue and aim to identify sufficient conditions ensuring the existence of an optimal noise function for a given distinguishability function $\ell(.)$, a given region $\calx$ and a loss function $\loss$. 
In our reasoning, we want $\ell(.)$, $\calx$ and also $\loss$ to be arbitrary. 
Therefore, instead of restricting the setting of these variables, we describe the conditions on the considered space $\Omega$ to have an optimal member. 

\paragraph{Conditions on the given space of noise functions.}
\begin{enumerate}
\item The first condition on $\Omega$ is that its members are \emph{uniformly bounded}. 
More precisely, there is some bound $M >0$ such that every $\calf \in \Omega$ satisfies $\calf(\vec v) \leq M$ almost everywhere in $\mathbb{E}^2$. 
This property is essential in certain cases to ensure that $\Omega$ admits an optimal member. 
For example, let the region $\calx$ be a single point in $\mathbb{R}^2$ and the loss function be increasing with the noise magnitude. 
In this situation, an optimal member of $\Omega$ can be obtained by assigning as much probability as possible to vectors having small magnitudes. 
Figure \ref{fig:opt-noise-bounds} illustrates this situation for various values of the bound $M$. 
It is clear from this figure that the optimal noise function (described by its radial $\calr(r)$) depends on $M$. 
However, if $M=\infty$ (\emph{i.e.}, $\Omega$ is not uniformly bounded), the expected loss is minimized by assigning probability $1$ to the zero-magnitude vector, but clearly in this case there is no noise (density) function. 
\begin{figure}
\centering
\includegraphics[width=0.50\textwidth]{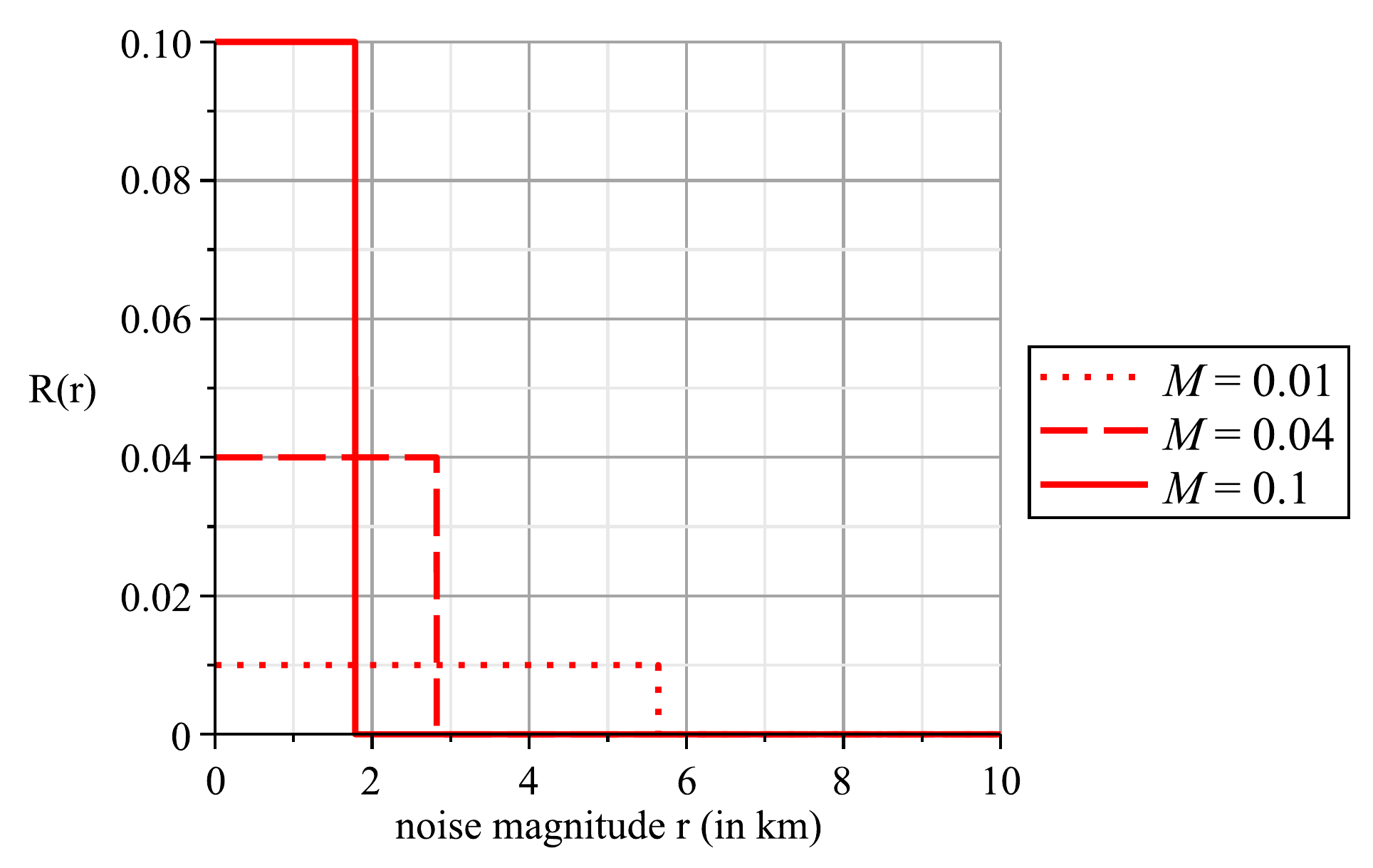}
\caption{The radials of optimal noise functions when $\calx$ consists of a single point.}
\label{fig:opt-noise-bounds}
\end{figure}
\item For any noise distribution $\calp$ and a noise magnitude $r$, let $\calp(>r)$ be the probability of sampling a noise vector for which the magnitude is larger than $r$. 
Based on the fact that the total probability assigned by $\calp$ to all noise vectors in $\mathbb{E}^2$ is $1$, it is intuitive that the probability $\calp(>r)$ converges to $0$ as $r \to \infty$. 
Using a function $\rho$ to precisely describe this convergence, we can parameterize this property on $\rho$, and say that $\calp$ is $\rho$-tight. 
More formally, we have the following.
\begin{definition}[$\rho$-tight noise distribution]\label{def:r-tight}
Consider a function $\rho:[0, \infty) \to [0,1]$ such that $\lim_{r\to \infty} \rho(r)=0$. 
A noise distribution $\calp$ is $\rho$-tight if for every noise magnitude  $r\geq 0$, it holds that $\calp(> r) \leq \rho(r)$.   
\end{definition}
Using the above property, we describe a second condition ensuring that $\Omega$ has an optimal member. 
More precisely, we require that there is a function $\rho$ such that all noise distributions of the members of $\Omega$ are \emph{uniformly} $\rho$-tight. 
This means that they have the same convergence rate (determined by $\rho$) for the probabilities of large noise\footnote{Uniform $\rho$-tightness of a collection of distributions is a stronger version of ``tightness'' (\emph{cf.}, page 59 in  \cite{billing:1999}), which is not parametric on $\rho$, and requires the probability masses to uniformly converge to zero outside any compact subset of $\mathbb{E}^2$.}.
\end{enumerate}

\paragraph{Conditions on the loss function.}
In addition to the above conditions on the considered space of noise functions, we also need a slight restriction on the loss function $\loss$. 
Precisely, it is required to be \emph{lower} semi-continuous at every $r_0 \in [0, \infty)$. 
This condition is written as $\liminf_{r \to r_0} \loss(r) \geq \loss(r_0)$, which means that in every neighborhood around $r_0$, the loss function $\loss$ has a minimum value. 
This condition is fundamental for the extreme value theorem that we use to prove that the expected loss attains its infimum in a space of noise functions. 
This condition is not too restrictive since it needs to be checked only at the discontinuities of $\loss$. 
In particular, it is enough to define the values of $\loss$ at every discontinuity $r_0$ to be $\liminf_{r \to r_0} \loss(r)$ to satisfy the lower semi-continuity. 

Based on the above conditions, we are now able to describe a space of noise functions that satisfy $\ell$-privacy on $\calx$. 
This space is defined by certain parameters, namely a bound $M>0$ and a function $\rho$, and therefore is written as $\Omega_{M,\, \rho}$. 
The following theorem shows that if $\Omega_{M,\, \rho}$ is non-empty, then it has an optimal member with respect to any 
lower semi-continuous loss $\loss$. 

\begin{theorem}[Existence of optimal noise functions]
\label{thm:opt-exists}
Consider a distinguishability function $\ell$, a set of locations $\calx$ and a lower semi-continuous loss function $\loss$. 
Consider also any $M>0$, and any $\rho:[0,\infty) \to [0,1]$ with $\lim_{r\to \infty} \rho(r) =0$. 
Let $\Omega_{M,\, \rho}$ be the space of all noise functions that are bounded by $M$ almost everywhere in $\mathbb{E}^2$, correspond to $\rho$-tight distributions and satisfy $\ell$-privacy on $\calx$. 
%
%
If $\Omega_{M,\, \rho}$ is non-empty, it has an optimal member for $\ell$, $\calx$ with respect to $\loss$. 
\end{theorem}
\begin{proof}
Let $\cald$ be the collection of every noise distribution that has a corresponding noise function (\emph{i.e.}, a pdf) in $\Omega_{M,\, \rho}$. 
Therefore, the expected loss is a real-valued function on the elements of $\cald$, and written as $\eloss(\calp, \loss)$ for every 
$\calp \in \cald$.  
Now, we proceed by showing that $\eloss(., \loss)$ attains a \emph{minimum} value in $\cald$. 

By the extreme value theorem, $\eloss(., \loss)$ attains a minimum in $\cald$ if the latter is non-empty, relatively compact and closed and finally $\eloss(., \loss)$ is lower bounded and lower semi-continuous in $\cald$. 
Since $\cald$ is non-empty (because $\Omega_{M,\, \rho}$ is), we prove in the following lines the other properties for $\cald$ and $\eloss(., \loss)$ using the weak convergence of probability distributions (known also as the convergence in distribution). 

\emph{Relative compactness:}
Since $\lim_{r \to \infty} \rho(r)=0$, for every $\sigma>0$, there is $r_\sigma$ such that $\rho(r_\sigma) < \sigma$. 
Since also every $\calp \in \cald$ is $\rho$-tight, we have by Definition \ref{def:r-tight} that $\calp(>r_\sigma) \leq \rho(r_\sigma)< \sigma$. 
Since the set of noise vectors having magnitudes $\leq r_\sigma$ is a compact subset of $\mathbb{E}^2$, it follows that $\cald$ is a tight collection of probability measures. 
Finally, by Prokhorov's theorem (\emph{cf.}, page 59 in \cite{billing:1999}), we conclude that $\cald$ is relatively compact. 

\emph{Closeness:} 
Whenever a sequence $\{\calp_n\}_{n\in\mathbb{N}}$ in $\cald$ converges weakly to $\calp$, which we write as $\calp_n \to \calp$, we want to show that $\calp \in \cald$. 
%
First, we show that $\calp$ has a density function. 
According to Portmanteau's theorem (\emph{cf.}~\cite[Theorem~1.3.4, p 18]{van:1996:weak}),
every open set $A\subseteq\mathbb{E}^2$ satisfies $\calp(A) \leq \liminf_{n \to \infty} \calp_n(A)$. 
Since for every $n \in \mathbb{N}$, the noise function $\calf_n$ of $\calp_n$ is bounded by $M$ almost everywhere in $\mathbb{E}^2$, we have $\calp(A) \leq\,M\,\lambda(A)$, in which $\lambda(A)$ is the area (\emph{i.e.}, Lebesgue measure) of $A$. 
More generally, we have 
\begin{equation}\label{eq:bound-on-prob}
\calp(B) \leq \calp(A) \leq M \,\lambda(A), \quad \forall B \subseteq A.
\end{equation} 
Using the fact that $\lambda(B)=\inf\{\lambda(A):\,A \supseteq B, A\, \mbox{open}\}$ (because $\lambda$ is an outer regular measure),  we get $\calp(B)=0$ when $\lambda(B)=0$. 
Therefore by the Radon-Nikodym theorem, there is a probability density function $\calf$ for the limit distribution $\calp$. 
%
We also show that $\calf$ is bounded by $M$ almost everywhere in $\mathbb{E}^2$.
Consider a ball $B_{\vec v} \subset \mathbb{E}^2$ around a vector $\vec v$. 
By the Lebesgue Differentiation Theorem,  $\calf(\vec v) = \lim_{\lambda(B_{\vec v}) \to 0} \calp(B_{\vec v})/\lambda(B_{\vec v})$ almost everywhere in $\mathbb{E}^2$. 
Since by Equation (\ref{eq:bound-on-prob}) $\calp(B_{\vec v}) \leq M\, \lambda(B_{\vec v})$ for every $B_{\vec v}$, we obtain 
$\calf(\vec v) \leq M$ almost everywhere in $\mathbb{E}^2$. 

%
Similarly, we show that $\calp$ is $\rho$-tight. 
It was proved that $\calp(B)=0$ when $\lambda(B)=0$ for all $B \subset \mathbb{E}^2$.   
Therefore $\calp$ assigns probability $0$ to the boundary of every set $V \subseteq\mathbb{E}^2$ (denoted by $\partial V$), and it holds by Portmanteau's theorem that  
%
\begin{equation}\label{eq:convergence-of-prob}  
\lim_{n\to\infty} \calp_n(V) = \calp(V),\quad \forall V \subseteq \mathbb{E}^2.
\end{equation} 
Since every $\calp_n$ is $\rho$-tight, $\calp_n(>r) \leq \rho(r)$ for all $r \geq 0$. 
Therefore, we imply by taking the limits and using Equation (\ref{eq:convergence-of-prob}) that $\calp$ satisfies this inequality too, 
and hence is $\rho$-tight.  
%
It remains to show that $\calp$ satisfies $\ell$-privacy on $\calx$.  
Since every $\calp_n$ satisfies $\ell$-privacy, we have by Equation (\ref{eq:l-privacy-prob}) 
\[
\calp_n(V) \leq e^{\ell(|\vec u|)} \calp_n(\tau_{\vec u}(V)) 
  \quad \forall V \subseteq \mathbb{E}^2, \forall \vec u \in \calv_\calx. 
\]
By applying the limits to the above inequality and using Equation (\ref{eq:convergence-of-prob}), we obtain $\calp(V) \leq e^{\ell(|\vec u|)}\, \calp(\tau_{\vec u}(V))$, which means that $\calp$ satisfies $\ell$-privacy on $\calx$. 
Thus $\cald$ is closed.  

\emph{Boundness from below and lower-semi-continuity:}
Let $Y$ be a random noise vector taking its values from $\mathbb{E}^2$. 
For all distributions $\calp_n \in \cald$, it is easy to see that $\eloss(\calp_n, \loss)$ is bounded from below by $0$ since it is the expected value (with respect to $\calp_n$) of $\loss(|Y|)$ that has this property. 
Since $\loss$ is lower semi-continuous on $[0,\infty)$, it is clear that $\loss(|.|)$ is also lower semi-continuous on $\mathbb{E}^2$. 
Thus, it follows by Portmanteau's theorem that every sequence $\calp_n \to \calp$ satisfies $\E_{\calp}[\loss(|Y|)] \leq \liminf_{n \to \infty} \E_{\calp_n}[\loss(|Y|)]$, which means that $\eloss(. , \loss)$ is lower semi-continuous on $\cald$.  
\end{proof}

Remark that for any $\ell$ and $\calx$, and any setting for the parameters $M, \rho$, there are always noise functions that are not members of the space $\Omega_{M,\, \rho}$, but yet satisfy $\ell$-privacy on $\calx$. 
In other words, $\Omega_{M,\, \rho}$ is restricted compared to the space of all noise functions that satisfy $\ell$-privacy. 
However, at the cost of this restriction $\Omega_{M,\, \rho}$ has the important feature that it admits an optimal member regardless of the choice of $\ell$ and $\calx$. 
Nevertheless, if specific assumptions are made on $\ell$, $\calx$, and the loss $\loss$, a space that is larger than $\Omega_{M,\, \rho}$ may also admit an optimal member. 
For example as shown in the following section, if the distinguishability takes the form $\ell(d)=\epsilon\,d$, the domain $\calx$ has a non-zero area, and the loss $\loss$ is increasing, the entire space of functions that satisfy $\ell$-privacy on $\calx$ has an optimal member. 

In conclusion, Theorem \ref{thm:opt-exists} opens a new avenue to explore the optimal noise functions, at least in the scope of $\Omega_{M,\, \rho}$, for various privacy requirements of the users on continuous regions. 
The analytical forms of such optimal noise functions depend indeed on the user-specific setting for $\ell$ and $\calx$ and $\loss$.   

%
%

\section{A case study: $\epsilon$-geo-indistinguishability}\label{sec:geo-ind}
The notion of $\epsilon$-geo-indistinguishability \cite{Andres:2013:indist} is an instance of $\ell$-privacy with the distinguishability function $\ell(d)=\epsilon\, d$. 
\mt{
In this setting, the parameter $\epsilon$ quantifies the allowed distinguishability for a unit distance, which corresponds to the maximum distinguishability between two points separated by one distance unit\footnote{
Since the distinguishability is unitless (as it is a ratio between two probabilities), the unit of $\epsilon$ is the reciprocal of the distance unit (\emph{e.g.}, $\textit{km}^{-1}$) and its numerical value depends indeed on the chosen unit for the distance.}.  
}

$\epsilon$-geo-indistinguishability models the situation in which the user requires to restrict the distinguishability between his location and every point at distance $d>0$ from him while enabling this restriction to be linearly relaxed as the distance $d$ increases. 

\subsection{Optimal noise function for $\epsilon$-geo-indistinguishability}
\label{sec:opt-geo-ind}

We consider an arbitrary geographical region $\calx$ that has a non-zero area, and aim to find the optimal noise function for $\calx$ with respect to $\epsilon$-geo-indistinguishability and an arbitrary increasing loss function $\loss$. 

In the sense of Definition \ref{def:opt} of optimal noise functions, we will implicitly assume $\Omega$ to be the entire space of noise functions satisfying $\epsilon$-geo-indistinguishability on $\calx$. 
This means that we require a noise function that minimizes the expected loss while satisfying $\epsilon$-geo-indistinguishability on $\calx$.  

The main tool to achieve this objective is Theorem \ref{thm:generality-circ} for which the assumptions are satisfied in the case of $\epsilon$-geo-indistinguish\-ability. 
More precisely, it is easy to see that  the distinguishability function $\ell(d) = \epsilon\, d$ is regular, and also satisfies $\ell(d_0 + d) \geq \ell(d_0) + \ell(d)$ for every $d_0>0$. 
Therefore, this theorem enables us first to focus only on circular noise functions and to reason about their radials. 
It also helps us to abstract from the geometry of the domain $\calx$ and focus on satisfying the privacy constraints on the entire space $\mathbb{R}^2$. 
Finally, we can assume without loss of generality that these constraints are \emph{strictly} satisfied for every pair of noise magnitudes in $[0, \infty)$.  

The first step towards achieving this objective is the following proposition that states that bounded and continuous circular noise 
functions are general enough to capture the required optimal noise function. 

\begin{proposition}[Bounded continuous circular functions are sufficient]
\label{prop:bounded-cont-geo}
Let $\calx$ be any region with a non-zero area.   
For every noise function $\calf$ satisfying $\epsilon$-geo-indisting\-uishability on $\calx$, and any loss function, there is a bounded and continuous circular noise function $\calf_\calr$ (with radial $\calr$) that has the same expected loss of $\calf$.
Furthermore, this function strictly satisfies $\epsilon$-geo-indisting\-uishability on $\mathbb{R}^2$ ,which means that
\begin{equation}\label{eq:geo-rad}
\calr(r)  \leq e^{\epsilon \, |r-r'|} \, \calr(r') \quad\quad \forall r, r' \in [0, \infty).    
\end{equation}
\end{proposition}
\begin{proof}
By Definition \ref{def:reg-distinguishability}, it is clear that $\ell(d)=\epsilon\,d$ is regular. 
In addition, since the area of $\calx$ is non-zero, it must contain an arbitrarily small disk with diameter $d_0>0$. 
It also holds that $\ell(d_0 + d) = \ell(d_0) + \ell(d)$ for all $d>0$. 
Thus, it follows from Theorem \ref{thm:generality-circ} that for every noise function $\calf$ satisfying $\epsilon$-geo-indistinguishability on $\calx$, there is a circular noise function $\calf_\calr$, with a radial $\calr$, providing the same expected loss of $\calf$. 
Furthermore, this function satisfies  $\epsilon$-geo-indistinguishability ``strictly'' on $\mathbb{R}^2$ in the sense of Theorem \ref{thm:circular-noise-suffice-l}. 
Observe that $\distset_{\mathbb{R}^2}=\{(r, r'):r,r' \in [0, \infty)\}$, and $\ell_{\mathbb{R}^2}(r,r') = \epsilon\, |r-r'|$. 
Therefore $\calr$ satisfies Inequality (\ref{eq:geo-rad}).   
%
%
Using this inequality, we show that $\calr$ is bounded as follows. 
By combining Inequality (\ref{eq:geo-rad}) and the total probability law \ref{eq:prob-law-radial}, 
we obtain  $\calr(r') \int_{r'}^\infty e^{-\epsilon(r-r')}\, 2 \pi r \,d r \leq 1$ for every $r'\geq 0$ . 
This yields that $\calr(r') \leq \epsilon^2 / 2\pi(1+\epsilon r') \leq \epsilon^2/2\pi$. 

Finally, to prove that $\calr$ is continuous everywhere in $[0, \infty)$, we consider any 
$r' \in [0, \infty)$ and show that $\lim_{r \to r'} |\calr(r) - \calr(r')|=0$. 
By Inequality (\ref{eq:geo-rad}), it is clear that $e^{-\epsilon |r-r'|}\, \calr(r') \leq \calr(r) \leq e^{\epsilon |r-r'|}\, \calr(r')$. 
Thus,
\begin{flalign*}
\calr(r) - \calr(r') \leq (e^{\epsilon |r-r'|} -1)\, \calr(r') \quad\mbox{if}\quad \calr(r)\geq \calr(r'), \\
\calr(r') - \calr(r) \leq (1- e^{-\epsilon |r-r'|})\, \calr(r') \quad\mbox{if}\quad \calr(r) <\calr(r'). 
\end{flalign*}
These two inequalities imply that 
\[
| \calr(r) - \calr(r') | \leq \max \{ (e^{\epsilon |r-r'|} -1),  (1- e^{-\epsilon |r-r'|}) \}\, \calr(r').
\] 
Taking the limits of the latter inequality when $r$ tends to $r'$ leads to $\lim_{r \to r'} |\calr(r) - \calr(r')|=0$. 
\end{proof}

Proposition \ref{prop:bounded-cont-geo} is an important outcome of the general analysis presented early in Sections \ref{sec:noise-distributions} and \ref{sec:cont-regions}. 
In particular, we used the results of that analysis to derive from the specification of the distinguishability function $\ell(d)= \epsilon\,d$ several analytical properties on the noise functions that are candidates to be optimal with respect to any loss function and any region having a non-zero area. 
In previous works \cite{ehab:2016:l-privacy}, these properties (specifically the continuity and the boundedness), were taken as assumptions limiting the range of considered noise functions.
In contrast, in our current analysis these properties are rather derived from the definition of the considered distinguishability function $\ell$. 

\begin{figure*}[]
\centering
\raisebox{0pt}{\includegraphics[width=0.55\textwidth]{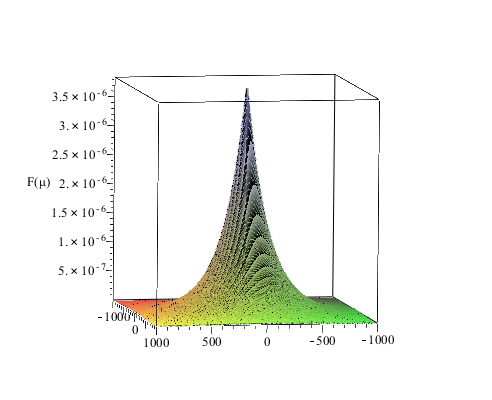}}
\hspace{-50pt}
\raisebox{20pt}{\includegraphics[width=0.40\textwidth]{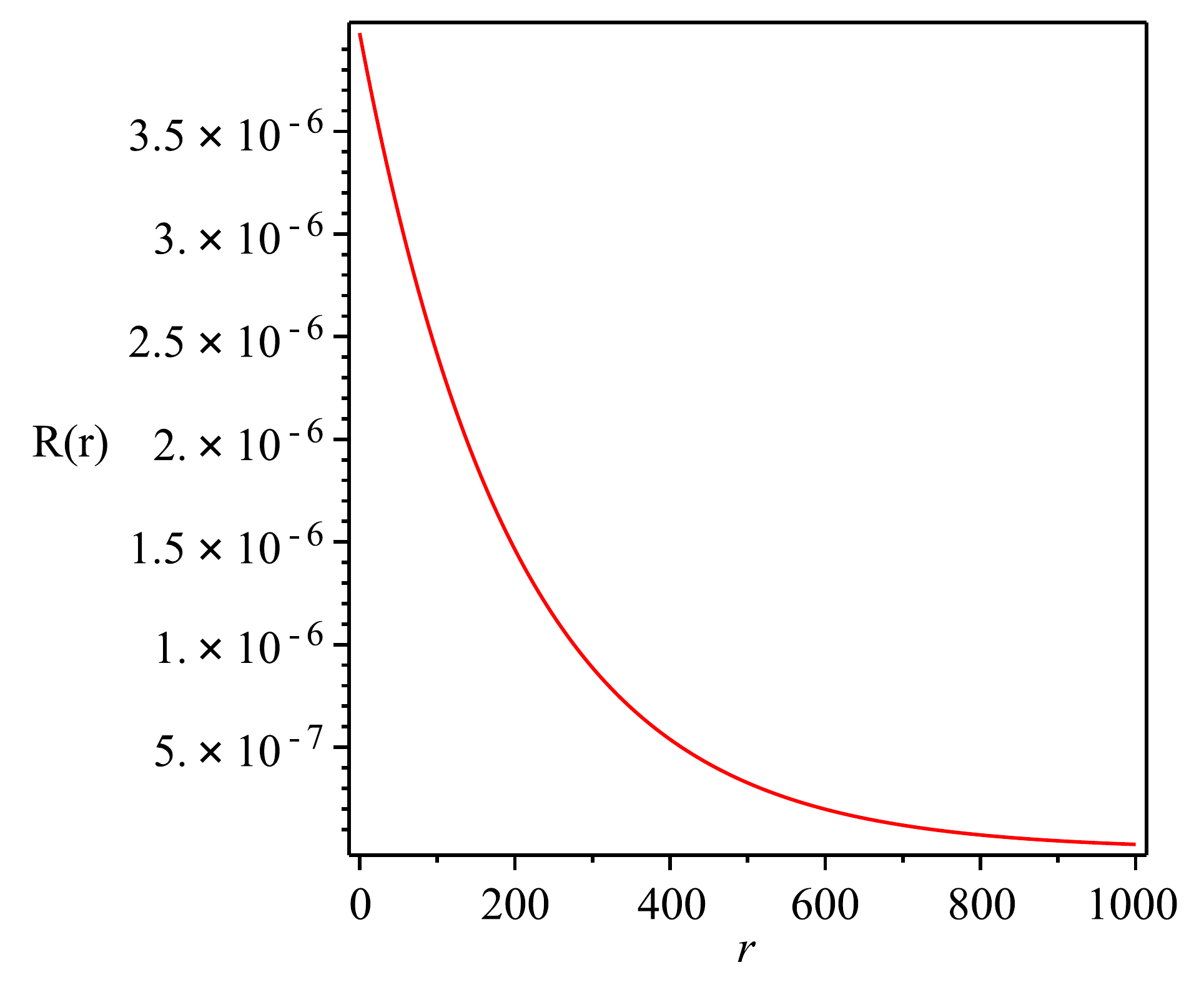}}
\caption{The planar Laplace noise function and its radial with $\epsilon=1/200$.}
\label{fig:lap-noise}
\end{figure*}

In the following, we go one step further by showing that a specific circular noise function, called the planar Laplace function, is optimal for \emph{every} geographical region having a non-zero area, and also with respect to \emph{every} increasing loss function. 
For a user-defined value of the privacy parameter $\epsilon>0$, the radial of this function decreases exponentially with the noise 
magnitude $r$ and precisely has the form $\calr(r) = \epsilon^2/(2\pi) \, e^{-\epsilon r}$. 
In Figure \ref{fig:lap-noise}, the Laplace noise function $\calf$ on the noise vectors $\mathbb{E}^2$ and its radial $\calr$ on the magnitudes of these vectors are illustrated for $\epsilon = 1/200$. 
While this function was originally introduced in \cite{Andres:2013:indist} as a candidate function to satisfy $\epsilon$-geo-indistinguishability, we show by the following theorem that it is furthermore optimal under the aforementioned conditions.     

\begin{theorem}[Optimality of the planar Laplace function for $\epsilon$-geo-indistinguishability]\label{thm:optimal-geo}
For any region $\calx$ having a non\-zero area, and any increasing loss function $\loss$, the Laplace noise function defined by the radial $\calr(r) = \epsilon^2/(2\pi) \, e^{-\epsilon r}$ with the parameter $\epsilon>0$ is optimal for $\calx$ with respect to $\epsilon$-geo-indistinguishability and $\loss$. 
\end{theorem}
\begin{proof}
According to Proposition \ref{prop:bounded-cont-geo}, it is sufficient to show that the Laplace function $\calf_\calr$ that has the radial $\calr(r) = \epsilon^2/(2\pi) \, e^{-\epsilon r}$ is optimal in the class $\calc$ consisting of every circular function strictly satisfying $\epsilon$-geo-indistinguish\-ability on $\mathbb{R}^2$, and has a continuous radial. 

First it is easy to verify that $\calf_\calr$ is a member of $\calc$ since its radial $\calr$ is clearly continuous everywhere in $[0, \infty)$ and satisfies Inequality (\ref{eq:geo-rad}). 
Thus, it remains to show that $\calf_\calr$ satisfies $\eloss(\calf_\calr, \loss) \leq \eloss(\calf_{\calr'}, \loss)$ for every other circular function $\calf_{\calr'}$ in $\calc$.  

Since both $\calr$ and $\calr'$ are continuous on $[0, \infty)$, their difference $g(r)=\calr(r) - \calr'(r)$ is also continuous on $[0, \infty)$. 
If $\calr, \calr'$ are not identical, there must be $r_1, r_2 \in [0, \infty)$ such that $g(r_1)>0$ and $g(r_2)<0$ because otherwise the total probability law \ref{eq:prob-law-radial} would not hold for either $\calr$ or $\calr'$. 
As a consequence by the intermediate value theorem, there must be $\bar r$ between $r_1$ and $r_2$, such that $g(\bar r)=0$, \emph{i.e.} $\calr(\bar r) = \calr'(\bar r)$. 
It also holds that $\calr(r) = \calr(\bar r)\, e^{-\epsilon(r-\bar r)}$ for all $r \in [0, \infty)$. 
Using these equalities along with the assumption that $\calr'$ satisfies Inequality (\ref{eq:geo-rad}), we can write 
\begin{flalign}
\forall r \leq \bar r: \quad   &\calr'(r) \leq \calr'(\bar r)\, e^{-\epsilon(r-\bar r)}=\calr(\bar r)\, e^{-\epsilon(r-\bar r)} 
= \calr(r), \label{eq:thm-geo-optimal-1} \\
\forall r > \bar r: \quad   &\calr'(r) \geq \calr'(\bar r)\, e^{-\epsilon(r-\bar r)}=\calr(\bar r)\, e^{-\epsilon(r-\bar r)}
= \calr(r). \label{eq:thm-geo-optimal-2}
\end{flalign}
We can also write
\[
\eloss(\calf_\calr, \loss) - \eloss(\calf_{\calr'}, \loss) = \int_0^\infty \loss(r) \left( \calr(r) - \calr'(r)\right)\, 2\pi r \, d r. 
\]
For all $r \in [0, \infty)$, it can be shown that $\loss(r) \left( \calr(r) - \calr'(r)\right) \leq \loss(\bar r) \left( \calr(r) - \calr'(r)\right)$ as follows. 
If $r\leq \bar r$ then $\loss(r)\leq\loss(\bar r)$ since $\loss$ is increasing, and $\calr(r) - \calr'(r) \geq 0$ by (\ref{eq:thm-geo-optimal-1}). I
f otherwise $r >\bar r $ then $\loss(r)\geq\loss(\bar r)$ and $\calr(r) - \calr'(r) \leq 0$ by (\ref{eq:thm-geo-optimal-2}).  
Thus, we conclude that
\[
\eloss(\calf_\calr, \loss) - \eloss(\calf_{\calr'}, \loss) \leq \loss(\bar r) \left(\int_0^\infty \left( \calr(r) - \calr'(r)\right)\, 2\pi r \, d r \right) = 0 
\]
in which the final equality follows from the fact that both $\calr$ and $\calr'$ satisfy the total probability law \ref{eq:prob-law-radial}. 
\end{proof}

The result stated by the above theorem is strong in two aspects. 
First, the Laplace noise function is optimal for every region having a non-zero area, regardless of the geometry and the size of the considered region. 
Furthermore, this optimality holds for all increasing loss functions, which are mostly used to quantify the loss of LBS quality due to the obfuscation. 
Thus, the user does not need to use a different noise function when he moves to a different region or when he uses a different 
loss function. 
Since Theorem \ref{thm:optimal-geo} describes the optimal noise function for $\epsilon$-geo-indistinguishability, it can be 
also interpreted in terms of the symmetric mechanisms presented earlier in Section \ref{sec:sym}. 
In particular, anyone of these mechanisms is based on a specific noise function used to sample the added noise vectors. 
Therefore Theorem \ref{thm:optimal-geo} identifies, under the stated conditions, the optimal symmetric mechanism satisfying 
$\epsilon$-geo-indistinguishability. 
In the following subsection, we compare between this mechanism and the instances of the other type of 
mechanisms, namely the non-symmetric ones.   

\subsection{Comparison to non-symmetric mechanisms on a coarse grid}\label{sec:comparison-coarse}

As described earlier, a symmetric mechanism has the characteristic that the probabilistic noise addition is independent of the user's location in the considered region $\calx$. 
In contrast, a non-symmetric mechanism samples the added noise using a noise distribution that depends on the real location of the user. 
One advantage of the latter approach is that it is more flexible. 
More specifically, the noise addition at every point of $\calx$ may be optimized using the user's prior in $\calx$ (\emph{i.e.}, the probability of the user to visit each point) such that the resulting mechanism has the minimum expected loss for the user while satisfying the privacy constraints. 
However it is clear in this case that the optimized non-symmetric mechanism depends on the considered region $\calx$, the user's prior $\vpi$ and the adopted loss function $\loss$. 
This means that if any of these parameters change, the user's device may need to compute another mechanism to query the LBS. 
In contrast, the symmetric mechanism that is based on the Laplace noise is optimal amongst symmetric mechanisms due to Theorem \ref{thm:optimal-geo}), and is insensitive to the changes of $\calx, \vpi$ and $\loss$. 

Another important issue is that constructing an optimal non-symmetric mechanism is infeasible when the considered region $\calx$ is continuous, because in this case the number of points in $\calx$ would be too large to apply the traditional linear optimization techniques. 
As shown by \cite{Bordenabe:2014:CCS}, such difficulty may be relaxed by discretizing $\calx$ into a \emph{coarse grid} having a small number of cells, and approximating $\calx$ by the centers of these cells. 
In this case, using a specific prior $\vpi$ on these centers, an optimal non-symmetric mechanism for $\vpi$ can be constructed.
While such a mechanism would by design satisfies indistinguishability constraints between those centers, it does not 
guarantee the indistinguishability between all points of the original continuous region $\calx$. 
Moreover, while this mechanism is optimal for a prior on the centers of the cells, it does not necessarily provide reasonable utility when the user is not at one of these centers. 
To compare the difference of utility provided by symmetric versus non-symmetric mechanisms, we experimentally compare in the following between the expected loss of a non-symmetric mechanism and the optimal symmetric one based on the Laplace noise. 

We consider a geographical region $\calx$ around the city of Los Angeles. 
This region is bordered by the latitudes 33.9301, 34.1996 and the longitudes -118.5354, -118.1010, which makes $\calx$ extending 30km south-to-north and 40km west-to-east. 
We then consider the symmetric mechanism that satisfies $\epsilon$-geo-indistinguishability on $\calx$ and is optimal with respect to the loss function $\loss(r)= r$ that grows linearly with the noise magnitude $r$. 
By Theorem \ref{thm:optimal-geo}, the noise function of this mechanism is the planar Laplace function equipped with the radial 
$\calr(r) = \epsilon^2/(2\pi) \, e^{-\epsilon r}$.
Thus, the expected loss of this mechanism is easily evaluated using Equation (\ref{eq:expected loss-radial}) to be $2/\epsilon$. 
For the comparison, we also construct a non-symmetric mechanism satisfying $\epsilon$-geo-indistinguish\-ability on $\calx$. 
To facilitate optimizing the expected loss of this mechanism for a given prior, we split $\calx$ into a coarse grid of $8\times6$ (\emph{i.e.}, 48) squared cells, approximate every location in $\calx$ by the center of the inclosing cell and restrict the output of the mechanism to be one of these centers. 
The required mechanism is then obtained by solving a linear program minimizing the expected loss for a given prior $\vpi$ on these centers, subject to $47\times48^2$ (\emph{i.e.}, 108 288) inequality constraints to satisfy  $\epsilon$-geo-indistinguishability, in addition to $48$ equalities.  
For every value of $\epsilon$ in the range $0.2$ to $3.0$ (with step size of 0.1), we construct this mechanism for four priors corresponding to four users. 
Each prior is precisely the probability distribution of the corresponding user to visit the individual 48 cells of $\calx$. 

\mt{
\paragraph{Construction of a prior using the Gowalla dataset.}
Gowalla is a geosocial network in which the users deliberately share their locations \cite{gowalla:2010}. 
These shared locations have been collected in the period from February 2009 to October 2010 to yield a dataset of 6 442 890 check-ins (locations) for 196 591 users. 
Every check-in is described by a record consisting of the user identifier, the latitude and longitude of his location and the time of checking-in. 
Using this dataset, we compute the prior of a specific user on our grid of Los Angeles by counting the number of his check-ins in every cell relative to his total number of check-ins in the entire grid. 
}

\paragraph{Results.}

\begin{figure*}[t]
\centering
\subfigure{\includegraphics[width=0.35\textwidth]{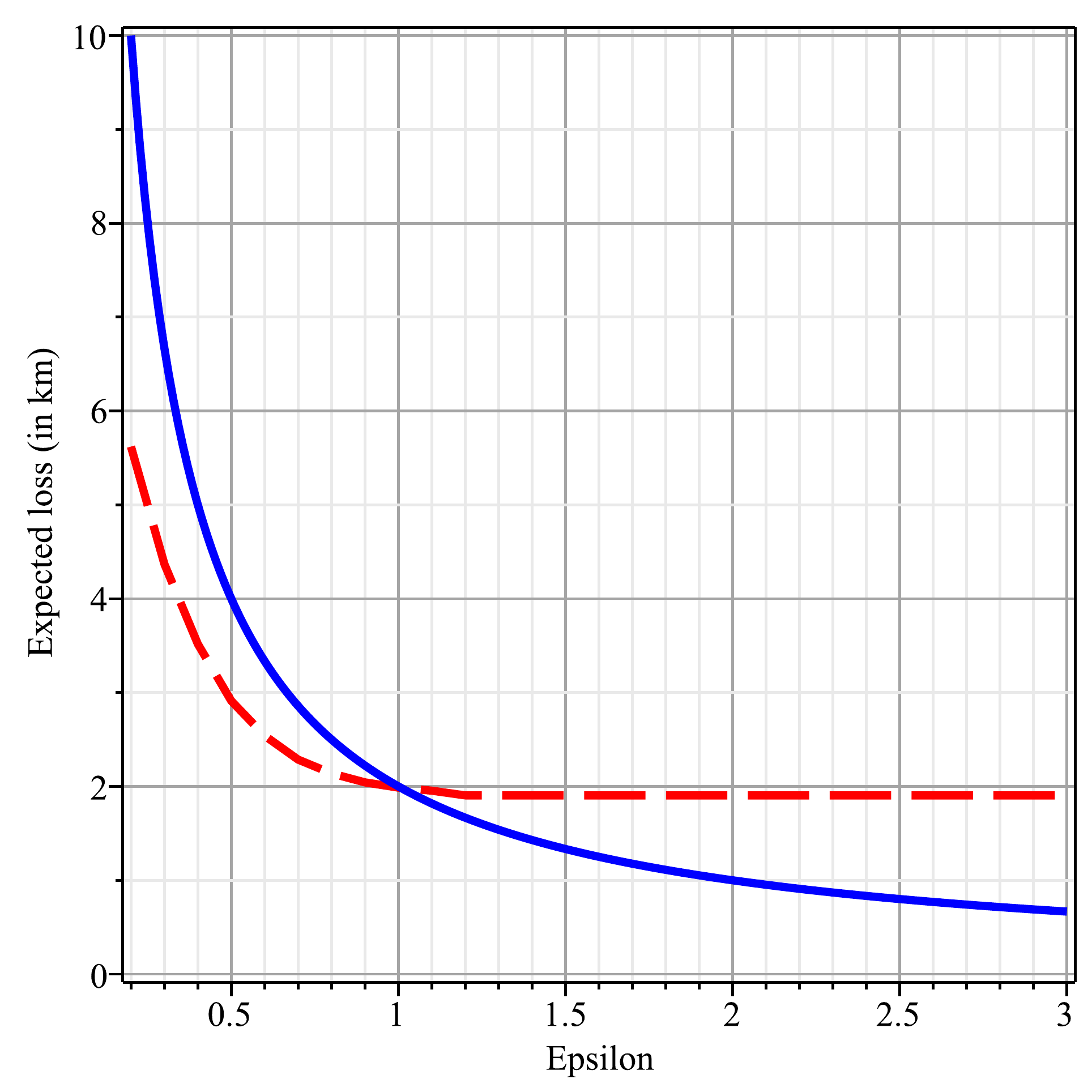}}
\subfigure{\includegraphics[width=0.35\textwidth]{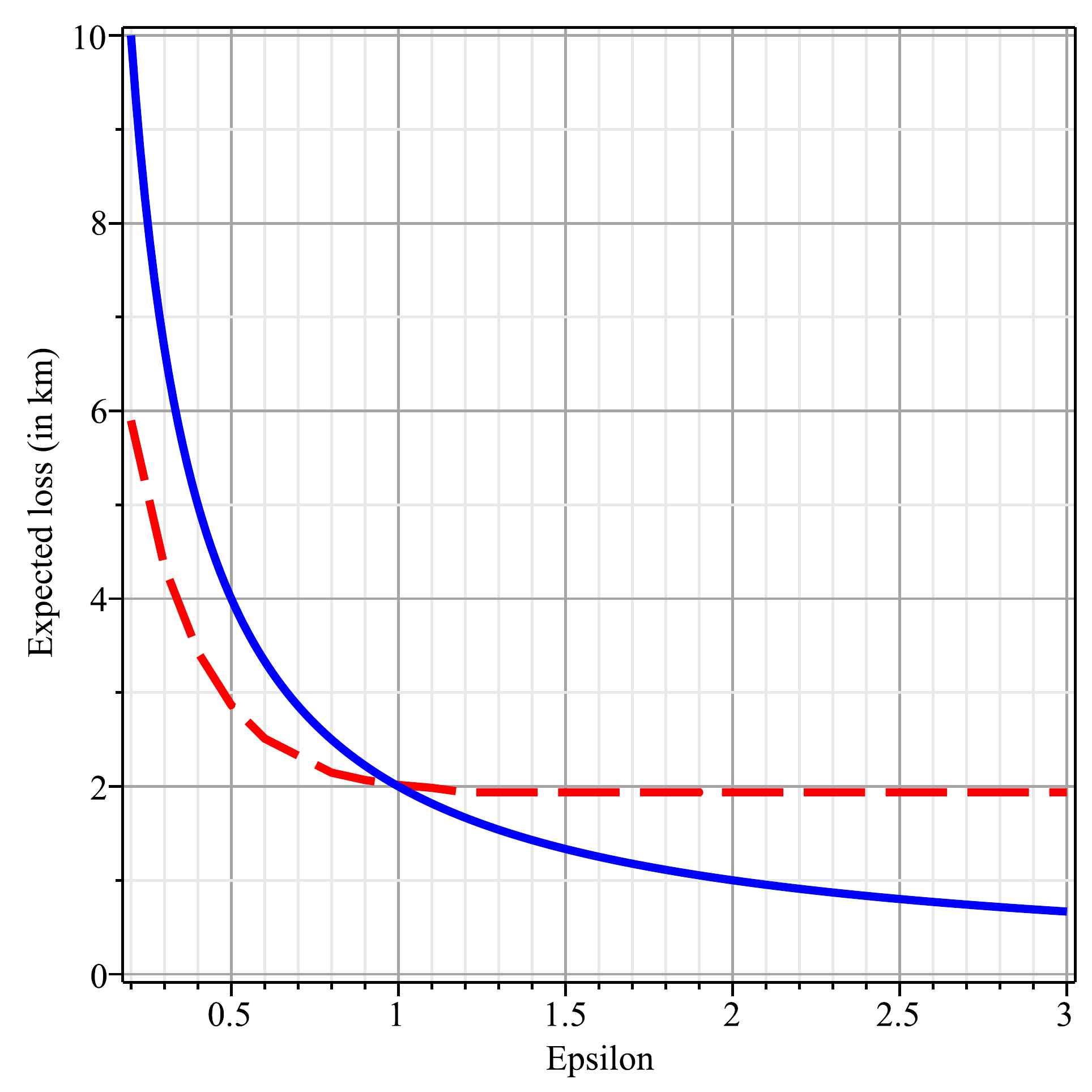}}
\subfigure{\includegraphics[width=0.35\textwidth]{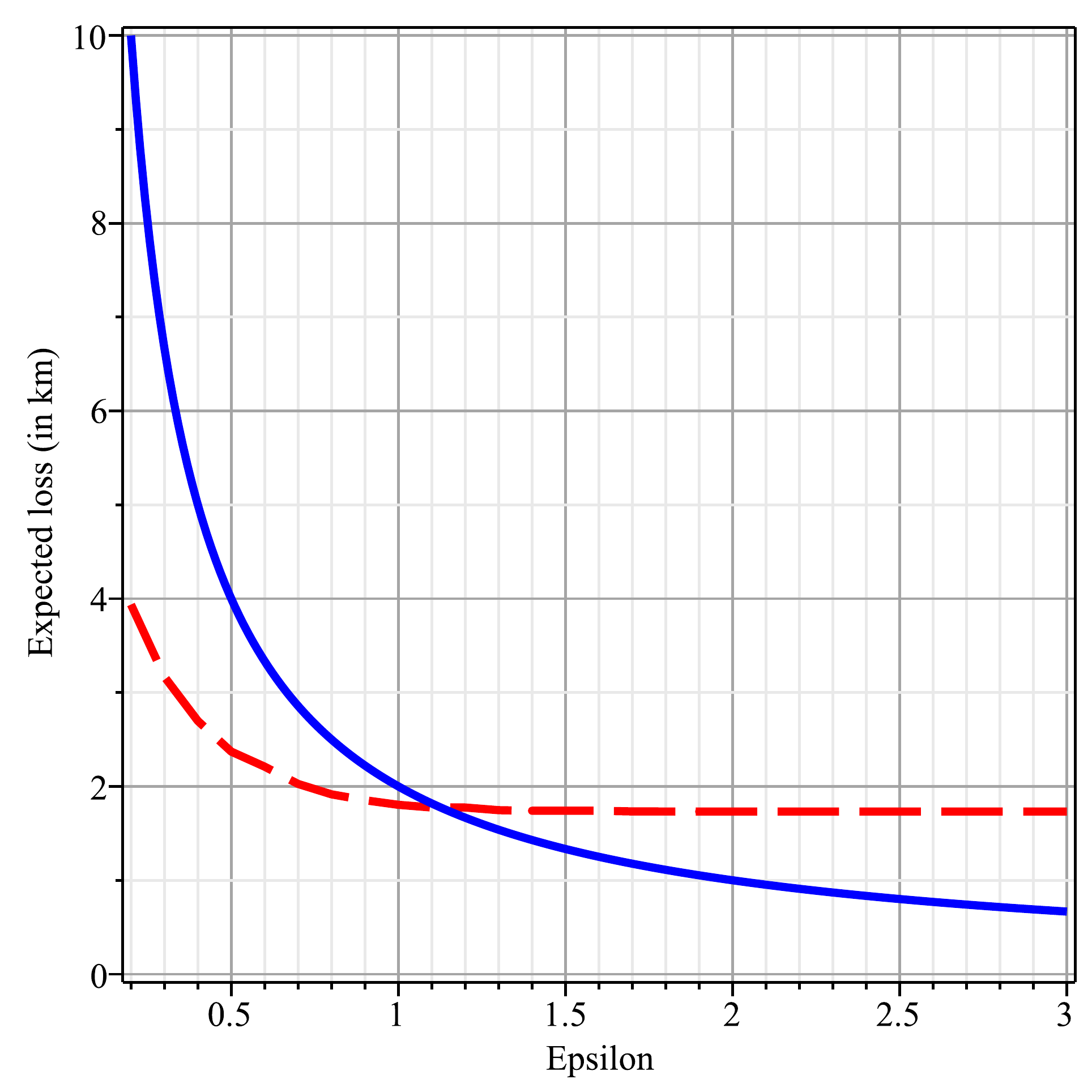}}
\subfigure{\includegraphics[width=0.35\textwidth]{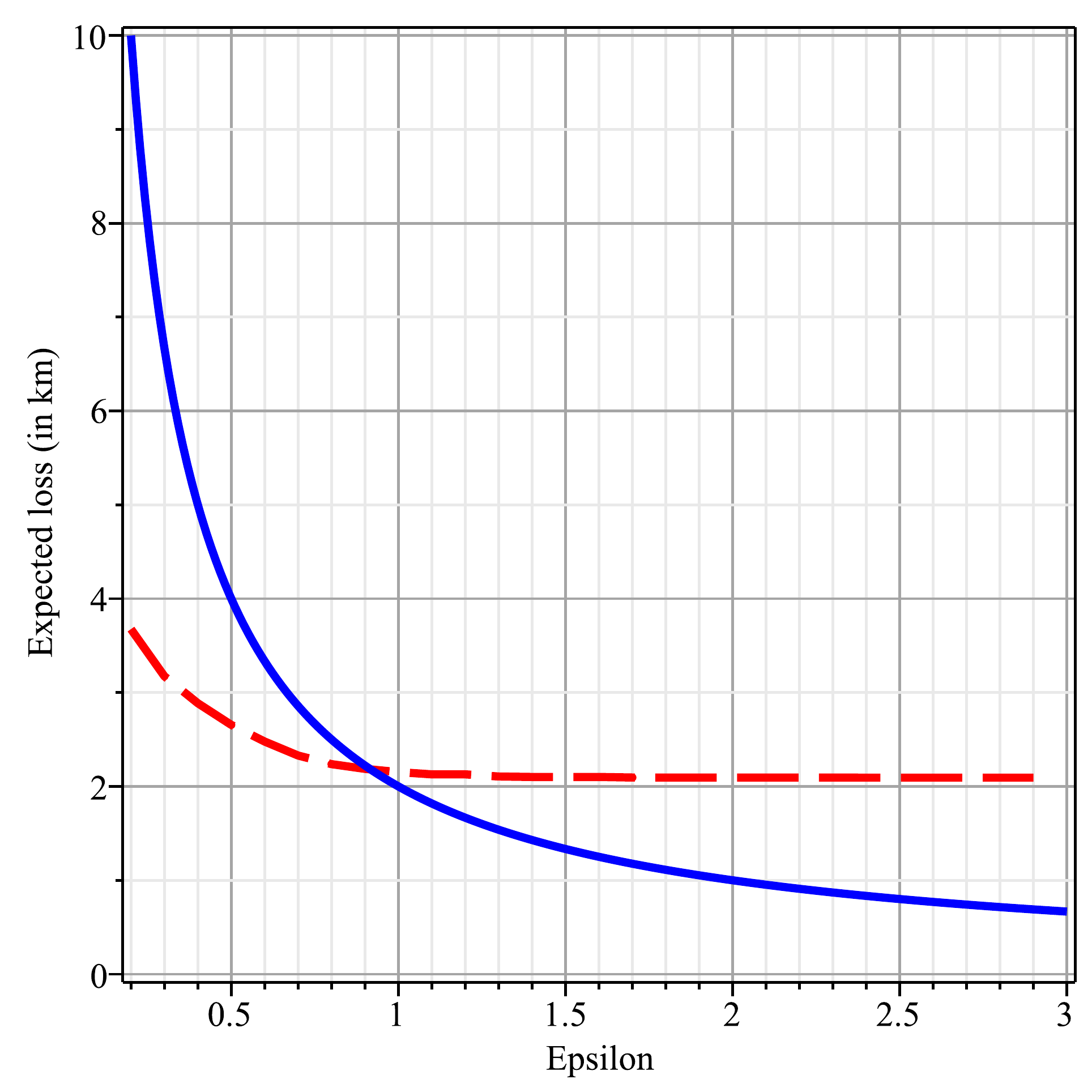}}
\caption{The expected loss (in kilometers) due to satisfying $\epsilon$-geo-indistinguishability for four users in Los Angeles. 
The solid curve corresponds to the symmetric mechanism with Laplace noise and the dashed curve corresponds to non-symmetric mechanisms personalized to every user.}
\label{fig:geo-ind-comparison}
\end{figure*}

Figure \ref{fig:geo-ind-comparison} plots the expected loss of the symmetric and non-symmetric mechanisms for the considered four users in Los Angeles. 
For each user, the solid curve corresponds to the symmetric mechanism, while the dashed curve corresponds to the non-symmetric one. 
%

From this figure, we can observed that the expected loss, for the two mechanisms, is non-increasing as $\epsilon$ grows. 
This is intuitive because as the privacy requirement modeled by $\epsilon$ is relaxed, one can always find a mechanism that 
has a better utility (\emph{i.e.}, lower expected loss). 
In particular, when the privacy is relatively strong (\emph{e.g.}, $\epsilon < 1$), the expected loss of both mechanisms is over 1.9 kilometer, and in this case the non-symmetric mechanism has a lower error compared to the other one. 
However, when the privacy level is more relaxed (\emph{e.g.}, $\epsilon>1$), the expected loss of the symmetric mechanism with the Laplace noise decreases at the rate $2/\epsilon$ to reach $0.66$km when $\epsilon$ is $3.0$, while the expected loss of the non-symmetric mechanisms tends to stabilize at a certain level around $2.0$km. 
This effect is due to the fact that the non-symmetric mechanism always maps the user's real location to the center of the enclosing cell, making the expected loss in the best case (\emph{i.e.}, with no obfuscation) be exactly the average distance between the user's real location and the center of his current cell. 
This level of saturation depends on the cell size and is expected to decrease as the region $\calx$ is fine-grained to smaller cells. 
However, optimizing the expected loss would be computationally more expensive in this case. 
As a conclusion, when the privacy is relatively strong, the level of expected loss for these two mechanisms is always high and in this case non-symmetric mechanisms may be favored. 
In contrast, when the required privacy level is more relaxed, the symmetric mechanism provides more reasonable levels of utility, compared to the non-symmetric mechanisms that are in all cases restricted by the limited computation resources. 

%
\mt{
\subsection{Remapping the outputs of a symmetric mechanism}

While a symmetric mechanism on the continuous region provides $\ell$-privacy for all points of the region, it may in some situations produce points that are unlikely to be visited by the user  
(\emph{e.g.}, inside a river or a sea). 
In this case, the output may be remapped to the nearest possible point (\emph{e.g.}, the side of the river or the sea). 
This remapping is a post-processing (of the output of the mechanism) that is independent of the original location of the user, and therefore preserves $\ell$-privacy as shown by \cite[Proposition~20]{ehab:2016:l-privacy}. 

A similar situation happens when the domain $\calx$ is a discrete set of points. 
In this situation, we can also use a symmetric mechanism to provide $\ell$-privacy for a continuous region covering 
these points, and remap its outputs to the discrete elements of $\calx$. 
In this case two techniques of remapping can be used. 
\begin{enumerate}
\item The output is remapped to its nearest element of $\calx$. 
\item Bayesian remapping can be applied in the same manner as in differential privacy \cite{Ghosh:09:STOC}. 
Using a prior distribution $\pi$ of visiting the points of $\calx$ and the mechanism $\calk$, 
a posterior distribution over $\calx$ is constructed after observing the output $z$. 
Afterwards $z$ is remapped to the point $R(z)$ that minimizes the expected loss with respect to the posterior distribution. 
More precisely 
\[
R(z) = \argmin_{z^*\in \calx} \sum_{x \in \calx} \pi(x)\,P(\calk(x)=z) \, \loss(x, z^*).
\]    
\end{enumerate}  
The above two techniques of remapping both yield two non-symmetric mechanisms satisfying $\ell$-privacy for the discrete region $\calx$ but they vary in terms of utility. 
To evaluate this aspect, we compare between the utilities of the two mechanisms in the case of $\epsilon$-geo-indistinguishability
as follows. 

We discretize the region of Los Angeles (described in Section \ref{sec:comparison-coarse}) into a fine grid of $80\times60$ cells in which the side length of every cell is 0.5 km, and construct the priors of two users using their check-ins in Gowalla dataset (they have 1120 and 753 check-ins in the region). 
Using the Laplace noise function, we construct the above (remapped) mechanisms for each user and evaluate their utilities for various values of $\epsilon$. 
Figure \ref{fig:geo-ind-comparison-lap-rem} demonstrates the results of this experiment, which shows that the Bayesian remapping is significantly better than the other technique. 
This superiority is clearly due to the fact that Bayesian remapping is optimized to the user's prior unlike the other simple technique that is independent of it. 

\begin{figure*}[t]
\centering
\subfigure{\includegraphics[width=0.35\textwidth]{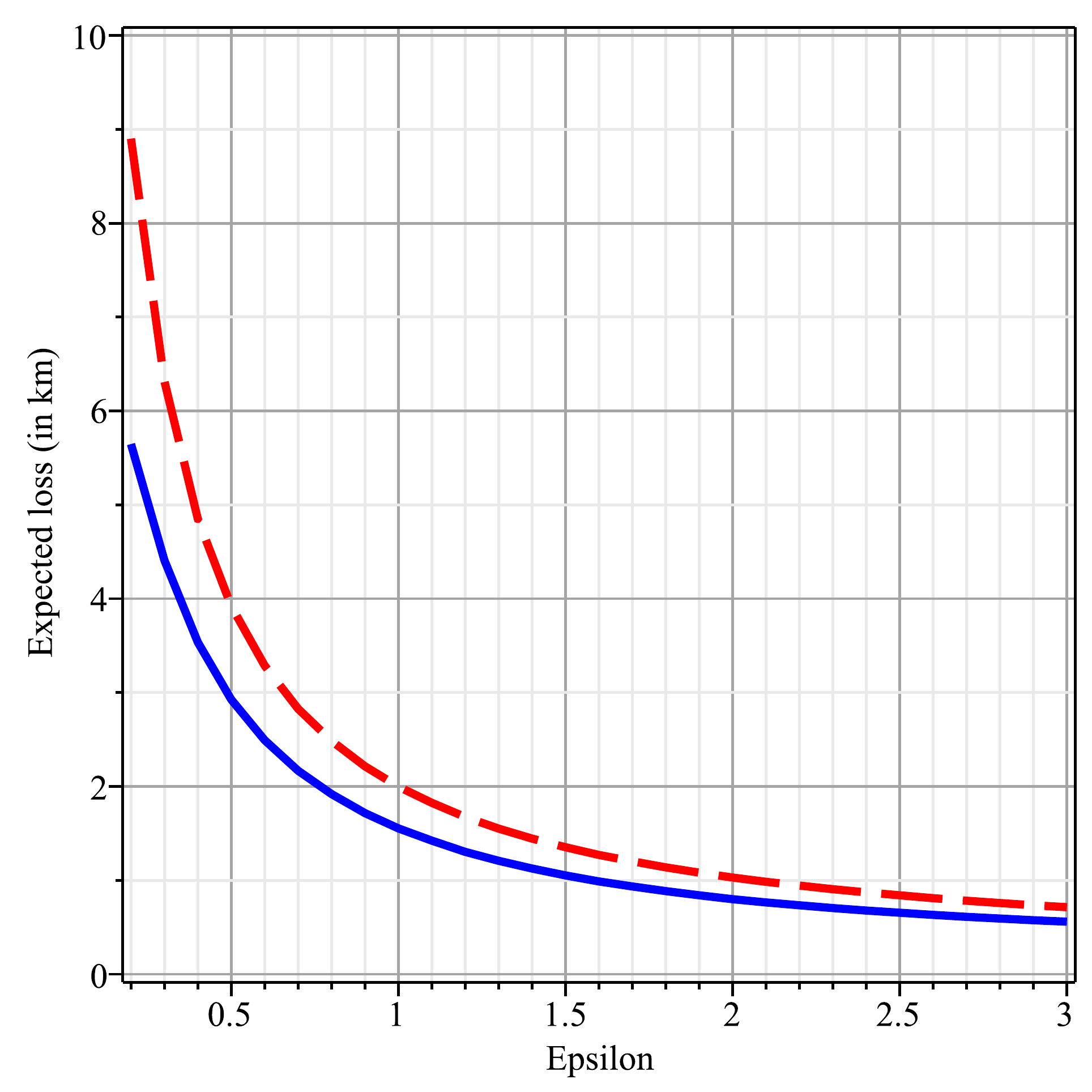}}
\subfigure{\includegraphics[width=0.35\textwidth]{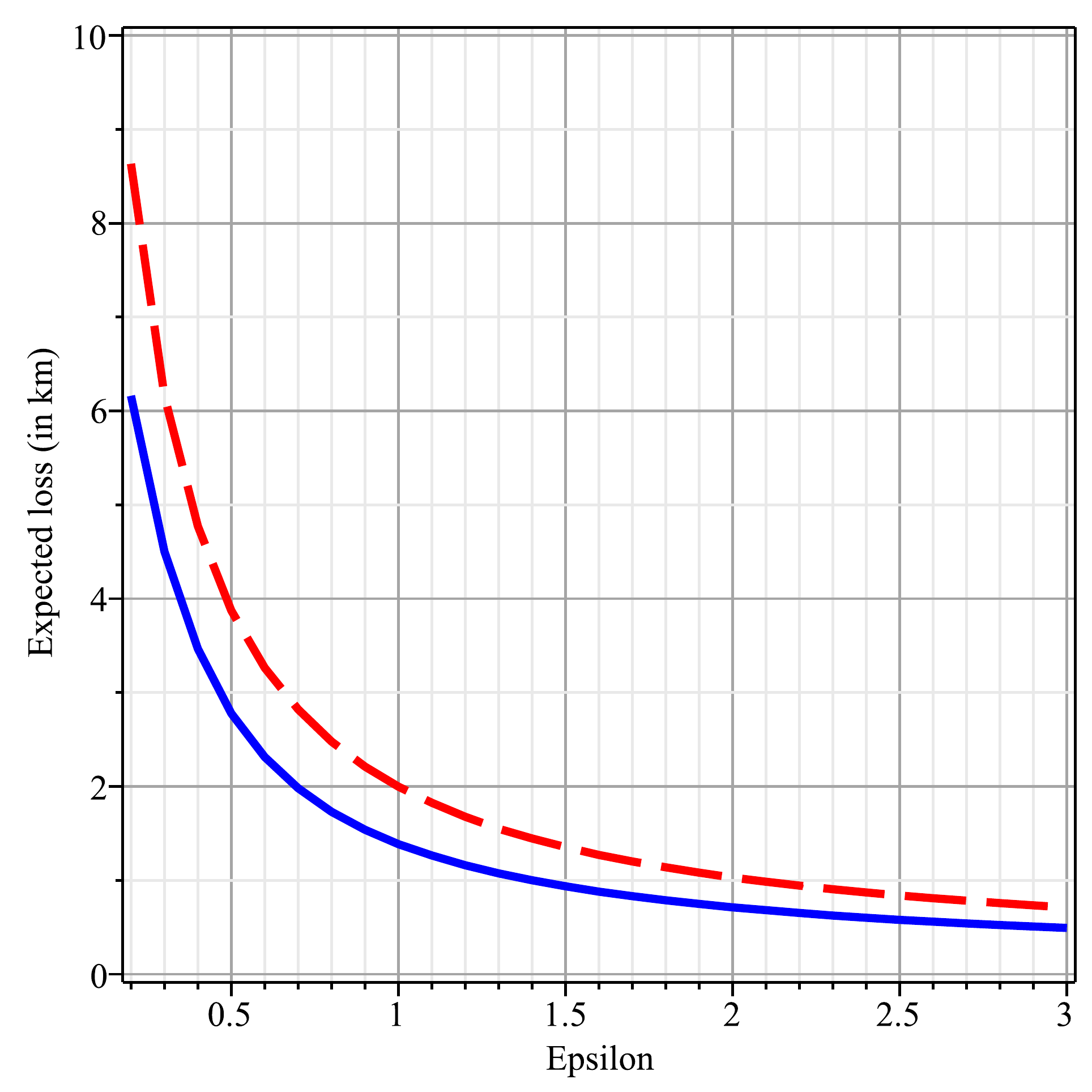}}
\caption{The expected loss (in km) of the mechanisms resulting from remapping Laplace mechanism for two users in Los Angeles.
The solid curve corresponds to Bayesian remapping of the output based on the user's prior while the dashed curve corresponds 
to remapping the output to the nearest point in the grid.} 
\label{fig:geo-ind-comparison-lap-rem}
\end{figure*}
}

%
%
\section{Conclusion}
\label{sec:conclusions}

The main objective of our work was to optimize the utility of the mechanisms accessing LBSs while satisfying a certain level of location privacy for their users. 
More precisely, we considered mechanisms that obfuscate the user's location before querying the LBS such that certain privacy requirements are satisfied while at the same time minimizing the degradation of the service utility due to this obfuscation. 

We model the user's location privacy generically by $\ell$-privacy \cite{ehab:2016:l-privacy} in which the privacy requirements are precisely described by the distinguishability function $\ell(.)$. 
This notion is an adaptation of differential privacy \cite{Dwork:06:ICALP} that restricts the distinguishability between every two 
databases differing in the data of one participant, hence protecting the privacy of participants in these databases. 
Based on the discrete characteristic of the query results of databases, linear optimization techniques have been used to 
construct ``optimal'' mechanisms to query them while satisfying differential privacy 
\cite{Ghosh:09:STOC,Gupte:10:PODS,Brenner:10:FOCS,ElSalamouny:13:POST,ElSalamouny2014}. 
However, we have shown that these techniques are not practical to construct an optimal privacy mechanism that satisfies 
$\ell$-privacy on continuous regions. 
Therefore, we chose to focus on ``symmetric'' mechanisms that satisfy $\ell$-privacy for a user by adding to his location 
a noise vector that is sampled according to a noise distribution $\calp$ on the vector space $\mathbb{E}^2$.  
We described the conditions on $\calp$ and its corresponding noise function to achieve $\ell$-privacy on a given geographical region $\calx$. 
In addition, when $\calx$ has a non-zero area and satisfies, together with the distinguishability function 
$\ell(.)$, a certain condition we proved by Theorem \ref{thm:noise-dist-cont-region} that satisfying 
$\ell$-privacy on $\calx$ is equivalent to satisfying it on the entire space $\mathbb{R}^2$. 
This results implies that the optimal noise function for $\mathbb{R}^2$ is also optimal for every region satisfying the 
condition stated in the above theorem, making it unnecessary to change the noise function when the user 
moves to a different region. 
Furthermore Theorem \ref{thm:generality-circ} strengthens this result and confines the choice of such optimal noise function to the class of circular noise functions.  
Since optimal noise functions do not always exist for given $\ell(.)$ and $\calx$, we described a parametric space $\Omega_{M, \rho}$ of noise functions, and proved by Theorem \ref{thm:opt-exists} that this space has always an optimal member regardless of $\ell$ and $\calx$. 

Finally as a special case of $\ell$-privacy, we considered $\epsilon$-geo-indis\-tinguishability and derived for it an optimal noise function.
More precisely we show by Theorem \ref{thm:optimal-geo} that the planar Lap\-lace function is optimal for \emph{every} region with a non-zero area and \emph{every} increasing loss function. 
Finally, we compared between the utility of the symmetric mechanism that uses this function to draw the added noise vectors, and the non-symmetric one constructed using the linear optimization 
techniques as in \cite{Bordenabe:2014:CCS}. 
To achieve a reasonable level of expected loss, the privacy level has to be relaxed, and in this case it was seen that the discretization error of the non-symmetric mechanism becomes significant compared to symmetric one, making the latter more favored. 

As future work, we plan to consider other instances of $\ell$-privacy, \emph{e.g.} ($D$, $\epsilon$)-location privacy and more generally the class of $D$-restricted distinguishability functions. 
We believe that the framework that we have introduced provides the basic tools to identify the optimal noise functions for these instances.    








\end{document}